\documentclass[showpacs,amsmath,amssymb,twocolumn,nofootinbib]{revtex4}
\usepackage{color}
\begin{document}
\allowdisplaybreaks[1]
\title{One-loop divergences for gravity non-minimally coupled to a multiplet of scalar fields:
calculation in the Jordan frame. I. The main results.}

\author{Christian F.~Steinwachs}
\affiliation{Institut f\"ur Theoretische Physik, Universit\"at zu K\"oln, Z\"ulpicher Stra\ss e 77, 50937 Cologne, Germany}
\author{Alexander Yu.~Kamenshchik}
\affiliation{Dipartimento di Fisica and INFN, Via 
Irnerio 46, 40126 Bologna, Italy\\
L.D. Landau Institute for Theoretical Physics of the 
Russian Academy of Sciences, Kosygin str. 2, 
119334 Moscow, Russia}
\begin{abstract}
Using the generalized Schwinger-DeWitt technique, we calculate the divergent part of the one-loop effective action for gravity non-minimally coupled to a multiplet of scalar fields. All the calculations are consistently done in the Jordan frame. 
\end{abstract}
\maketitle 
\section{Introduction}
One of the main problems in quantum field theory is the appearance of ultraviolet 
divergences and the necessity of their renormalization (see e.g.~\cite{BSH,IZ,W}). 
However, the quantum field divergences are not only the cause of troubles, but can also be 
an important source of information. The requirement of renormalization invariance, i.e. the independence of the physically measurable quantities of the parametrization of 
the renormalization procedure, requires that the renormalization group equations 
\cite{Stuch,Gell-Mann,BSH} are fulfilled. 
Their solutions permit in turn to connect the values of the physical quantities under study at different energy scales.
The theory of renormalization works very well for renormalizable theories, where all the divergences can 
be eliminated by means of introducing a finite number of counterterms to the Lagrangian. But its application becomes 
problematic in the case of non-renormalizable theories, where the number of divergent structures is unlimited. 
As it is well known, quantum gravity is a non-renormalizable theory \cite{BD,W1,QG}. Indeed, the gravitational coupling constant 
has mass dimension minus two. Thus, the Feynman diagrams, which contain a growing number of graviton loops, lead formally to an 
infinite set of different counterterms. As regards pure gravity theory, it was shown that at the one-loop level no physically relevant divergences remain on mass shell; all of them can be absorbed by a field redefinition \cite{HV}.
However, at the two-loop level, pure gravity is non-renormalizable \cite{GS}.
If gravity interacts with matter, namely with a scalar field, it becomes non-renormalizable already in the one-loop approximation
\cite{HV}. For the situation of renormalizability in the presence of SUSY see e.g.~\cite{SUSYGravity}.

Nevertheless, from the effective field theory point of view \cite{W} it makes sense to work with the usual non-renormalizable Einstein gravity and to apply such a useful tool 
as the renormalization group equations. Weinberg has suggested that concepts like asymptotic safety could be 
very fruitful in the application of the renormalization group to quantum gravity \cite{W1}. A theory is considered to be asymptotically safe 
if ``essential coupling parameters'' approach a non-trivial fixed point as the momentum scale goes to infinity. Asymptotic safety can be 
treated as a generalization of the notion of renormalizability, which fixes all but a finite number of coupling constants of a 
theory. Nowadays, the study of asymptotic safety and fixed points in quantum gravity has undergone an immense development 
\cite{W2,Percacci,Litim,Reuter}.

It is also worth noticing that the standard formalism of the renormalization group can be generalized to the case of non-renormalizable theories \cite{Kazakov}. This procedure looks especially natural in the framework of dimensional regularization \cite{dimen}.

The study of the interaction of gravity with matter is important from both the theoretical and phenomenological 
point of view. First of all, let us notice that considering the Einstein theory in the presence of a scalar field, it is natural 
not to limit ourselves by just introducing the metric into the kinetic term for the scalar field (the so-called minimal interaction),
but to accept the modification of the Einstein-Hilbert curvature term by adding a term proportional to $R\phi^2$, where 
$R$ is the scalar curvature and $\phi$ is a scalar field. Indeed, such a structure arises  in a natural way as a one-loop quantum correction to 
the Lagrangian of the scalar field in a curved spacetime, a fact that has stimulated the elaboration of the theory of induced 
gravity \cite{ind,ind1}. Furthermore, it was noticed that cosmological models with non-minimal coupling between an inflaton scalar field and 
gravity have some advantages compared to inflationary models based on a minimally coupled inflaton scalar field 
\cite{Spok,Fakir,Salop,Komatsu}.
The models with non-minimal coupling were considered also in the framework of one-loop quantum cosmology \cite{we,we1,we2}
to study the properties of the no-boundary \cite{HH,HH1} and tunneling \cite{tun,tun1,tun2,tun3}
wave functions of the universe.

While old cosmological models with a non-minimally coupled scalar field \cite {Spok}-\cite{we2}  have explicitly or implicitly associated such a field 
with Grand Unification Theories, recently the idea was put forward that the non-minimally coupled inflaton scalar field is the Standard Model Higgs boson \cite{Shap}. In \cite{BKS} it was shown that quantum effects are essential for the 
correct description of the inflationary dynamics in the model based on the non-minimally coupled Higgs field. 
In \cite{Shap1,Shap2,BKKSS,BKKSS1,Wil} the renormalization group formalism was used to establish a relation between 
the known values of the coupling constants of the Standard Model at the electroweak scale to their values at the 
inflationary scale in the presence of a strong non-minimal coupling between the Higgs field and gravity.   
In \cite{Clark,Lerner} the same problem was considered with an additional singlet scalar field that can be responsible for the presence of dark matter. The fitting of the effective Higgs inflationary potential in the model with non-minimal coupling  directly to WMAP5+BAO+SN data was undertaken in \cite{Popa}. In \cite{seesaw} the non-minimal inflation scenario was combined with the seesaw mechanism, 
whereas in \cite{supergravity} the relation between non-minimal inflation and supergravity was studied.

The renormalization group calculations performed in \cite{Shap1,Shap2,BKKSS,BKKSS1,Wil} give qualitatively similar results, which 
suggests that the non-minimally coupled Higgs inflation model is compatible with observations. 
However, there are some quantitative differences between the results obtained in these papers. The roots of the discrepancies lie 
in the different treatment of the suppression mechanism for the contribution of gravitons and the Higgs scalar field mode to the quantum loops.
Thus, the study of such subtle questions as the relation between Jordan and Einstein frames (see e.g.~\cite{Lerner1,BMSS}), 
the parametrization of the scalar field multiplets, and the conservation of the gauge invariance becomes urgent. 
Moreover, the fact that we do not consider a single scalar field, but a multiplet of scalar fields, causes the transformation from the Jordan frame to the  Einstein frame and back to be more involved \cite{Hertzberg,Kaiser}. 

In the present paper we shall calculate the one-loop divergences in the theory with an arbitrary 
non-minimal coupling between the multiplet of scalar fields and gravity in the \emph{Jordan frame}. Here it is necessary to 
emphasize that working with the non-minimally coupled scalar fields, one often uses both frames simultaneously. In 
\cite{we-renorm} the one-loop divergences for a singlet scalar field non-minimally coupled to gravity were calculated for 
a system of generalized potentials, and the formalism for the generalized renormalization group equations was developed.
Starting from the action written in the Jordan frame, the authors of \cite{we-renorm} have undertaken the transition to the Einstein frame by a conformal 
transformation of the metric and by a proper redefinition of the scalar field such that the interaction between the scalar field 
and gravity becomes minimal and the kinetic term for the scalar field acquires the standard form. 
Then the one-loop divergences were calculated by using the generalized Schwinger-DeWitt technique \cite{DeWitt,Bar-Vil}
for the fields in the Einstein frame. Finally, the result was rewritten in terms of the original Jordan frame fields.
The conformal transformation between the Jordan and Einstein frames is a legitimate operation at the classical level, 
but it becomes dangerous when the quantization of fields is involved. 

For this reason, in the present paper we perform all calculations exclusively 
in the Jordan frame and present the divergent part of the one-loop corrections as our main result. In the subsequent paper \cite{we-future},
these results are used to compare them with the results obtained by the methods of \cite{we-renorm} for the case of a multiplet of scalar fields. The third paper \cite{we-future1} of the series will be devoted to important applications within cosmological models based on a non-minimally coupled Higgs field that plays the role of the inflaton.

The structure of the paper is as follows: in Sec.~II we briefly describe the algorithm of the calculations for the divergent part of the one-loop effective action. In Sec.~III we give the complete expression for this divergent part. Sec.~IV is devoted to cross-checks and comparison with known results. A major source for such a comparison are the results of Shapiro and Takata \cite{shapiro}, who have made similar calculations for the case of a single scalar field. Sec.~V contains concluding remarks and the announcement of the content of the subsequent papers.

\section{One-loop divergences in the Jordan frame: the algorithm}
We consider a model with the action
 \begin{align}
S = \int_{{\cal M}} \text{d}^4x\sqrt{g}\left(U(\varphi)R - \frac12 g^{\mu\nu}G(\varphi)\nabla_{\mu}\Phi^{a}\nabla_{\nu}\Phi_{a}-V(\varphi)\right),  
\label{action} 
\end{align}
where 
\begin{align}
\varphi \equiv \sqrt{\delta_{ab}\Phi^a\Phi^b},\  a = 1,\cdots,N\;,
\label{varphi}
\end{align}
that is, gravity interacts with a multiplet of real scalar fields.
The generalized potentials $U,G$ and $V$ are invariant with respect to rotations in the 
$N$-dimensional space and are thus ultra-local functions of the modulus $\varphi$. It is convenient to consider a Euclidean manifold ${\cal M}$. 
We use the following definition of the Riemann tensor:
\begin{align}
R^{\alpha}_{\ \mu\nu\beta} = \frac{\partial \Gamma^{\alpha}
_{ \mu\beta}}{\partial x^{\nu}}-\frac{\partial \Gamma^{\alpha}_{\mu\nu}}{\partial x^{\beta}}  + 
\Gamma^{\lambda}_{\mu\beta}
\Gamma^{\alpha}_{\nu\lambda}-
\Gamma^{\gamma}_{\mu\nu}\Gamma^{\alpha}_{\beta\gamma}\;.
\label{Riemann}
\end{align}
In what follows, we use the background field method and split the generalized field
\begin{align}
\psi_{A} = \left(\begin{array}{c}
g_{\alpha\beta}\\
\Phi_a
                \end{array}\right)
\label{psi}
\end{align}
into a background part $\bar{\psi}_{A}$ and fluctuations $\delta\psi_{A}$, where
\begin{align}
\bar{\psi}_{A} = \left(\begin{array}{c}
\bar{g}_{\alpha\beta}\\
\bar{\Phi}_a
                \end{array}\right)\;,
\label{psi0}
\end{align}
and
\begin{align}
\delta\psi_{A} = \left(\begin{array}{c}
\delta g_{\alpha\beta}\\
\delta\Phi_a
                \end{array}\right) \equiv 
\left(\begin{array}{c}
h_{\alpha\beta}\\
\sigma_a
                \end{array}\right).
\label{psi1}
\end{align}
The action (\ref{action}) should be complemented by the gauge-breaking term
\begin{align}
S_{\text{GB}} = -\frac12 \int_{\cal M} \text{d}^{4}x\, \sqrt{g}\; \chi_{\mu}\chi^{\mu},
\label{GB}
\end{align}
where the function $\chi_{\mu}$ represents the generalization of the well-known background covariant DeWitt condition 
\cite{DeWitt,Bar-Vil}:
\begin{align}
&\chi_{\mu} = \sqrt{U}\left(\nabla^{\alpha}h_{\alpha\mu} -\frac12\nabla_{\mu}\delta h + 
f^{a}(\varphi)\nabla_{\mu}\sigma_{a}\right)\;,
\label{GB1}
\end{align}
where $h\equiv g^{\alpha\beta}h_{\alpha\beta}\;.$ 
Here, $f^{a}(\varphi)$ is an arbitrary function whose explicit form will be chosen later.

To calculate the one-loop divergent part of the effective action in the model with the action (\ref{action}) 
using the background field formalism \cite{DeWitt,Bar-Vil}, we need to know the second-order differential operator 
\begin{align}
F_{AB} = \frac{\delta^2 (S + S_{\text{GB}})}{\delta \psi_{A} \delta \psi_B}\;.
\label{second-order}
\end{align}
We also have to take into account the contribution of the Faddeev-Popov ghost term, which is given by the determinant of the ghost operator $Q^{\alpha}_{\beta}$ defined as 
\begin{align}
Q^{\alpha}_{\beta} = \frac{\delta{(\chi_{\xi})^{\alpha}}}{{\delta \xi^{\beta}}},
\label{FP}
\end{align}
where $\xi^{\beta}$ is a vector field realizing the general gauge transformation of the field variables by their Lie-dragging.

Let us now briefly remind the main steps of the generalized Schwinger-DeWitt algorithm for the calculation of the one-loop    
effective action \cite{DeWitt,Bar-Vil}.
The one-loop effective action for gauge theories has the following form:
	\begin{align}
	i\,W_{1-{\rm loop}}=&\,-\frac{1}{2}\,{\rm
Tr\;ln}\frac{\delta^{2}S_{\rm
	tot}[\psi]}{\delta
	\psi^{A}\delta\psi^{B}}+{\rm Tr\;ln}\,Q_{\alpha}^{\beta}\;.
	\end{align}
Here, $\psi^{A}$ is the full set of fields, $S_{\rm
tot}[\,\psi\,\,]$ is the total action
of the
theory including the gauge-breaking term and ${\rm Tr}$ is the
functional
trace. 
Generally, the second-order operator (\ref{second-order}) has the following form:
\begin{align}
 F_{AB} = C_{AB}^{\mu\nu}\nabla_{\mu}\nabla_{\nu} + 2\Gamma_{AB}^{\mu}\nabla_{\mu} + W_{AB},
\label{second-order1}
\end{align}
 where $\nabla_{\mu}$ is a covariant derivative defined with respect to some general affine connection.  

For the application of the Schwinger-DeWitt algorithm, the operator $F_{AB}$ should have the so-called minimal form, i.e. 
\begin{align}
 F_{A}^{\;B} = g^{\mu\nu}\,{\cal D}_{\mu}{\cal D}_{\nu}\, \delta_{A}^{\;B} + P_{A}^{\;B} - \frac{1}{6} R\,\delta_{A}^{\;B},
\label{minim0}
\end{align}
where ${\cal D}_{\mu}$ is a new covariant derivative defined with respect to a new connection and the term $R/6$ is subtracted for 
convenience. To arrive from (\ref{second-order1}) at the minimal form (\ref{minim0}) one should satisfy the condition
\begin{align}
C_{AB}^{\mu\nu} = \tilde{C}_{AB}\,g^{\mu\nu},
\label{minim}
\end{align}
and absorb the term $2\Gamma^{\mu}_{AB}\nabla_{\mu}$ linear in the derivative by defining the new covariant derivative ${\cal D}_{\mu}$.

The proper choice of the gauge condition (\ref{GB1}) guarantees that the condition (\ref{minim}) is satisfied. That means that 
the second-order derivatives in the operator $F_{AB}$ are proportional to the d'Alembertian. 
We then multiply the operator $F_{AB}$ by the inverse matrix $\tilde{C}^{(-1)CA}$ to eliminate the dependence of the operator 
(\ref{minim0}) on the matrix $\tilde{C}_{AB}$. Finally, 
in order to remove the term linear in the derivative we should redefine the affine connection by adding the term proportional to $\Gamma_{AB}^{\mu}$ to it .  

Now we are ready to apply the Schwinger-DeWitt algorithm. 
The logarithmically divergent part of the one-loop effective action is 
\begin{align}
	W_{1-\rm loop}^{\rm div}=&\,\frac{1}{32\pi^{2}(2-\omega)}\,
	\int \text{d}^{4}x\, 
	\sqrt{g}\,\text{tr}\,[(a_2)_{A}^{\;B}]\nonumber \\
&\,-\frac{1}{16\pi^{2}(2-\omega)} \,
	\int \text{d}^{4}x\,
	\sqrt{g}\,\text{tr}\, [(a_{2\,Q})_{\mu}^{\;\nu}]\;.
\label{div}	
\end{align}
Here, $(a_2)_{A}^{\;B}$ is the second coefficient of the Schwinger-DeWitt expansion for the operator $F_{A}^{\;B}$, whereas 
$(a_{2\,Q})_{\mu}^{\;\nu}$ is the analogous coefficient for the ghost operator $Q_{\mu}^{\;\nu}$. The parameter $\omega$ is the half-dimensionality of spacetime.
 
The general formula for the $\,(a_2)_{A}^{\:B}$ is \cite{Bar-Vil}
\begin{widetext}
\begin{align}
	(a_2)_{A}^{\;B}=\frac{1}{180}\left\{R_{\alpha\beta\mu\nu}
	^{2}-
	R_{\mu\nu}^{2}+\Box\vphantom{I}
	R\right\}\delta_{A}^{\;B}+
	\frac{1}{2}(P^2)_{A}^{\;B}+
	\frac{1}{12}({\cal R}_{\mu\nu}^2)_{A}^{\;B}+
	\frac{1}{6}(\Box\vphantom{I}P)_{A}^{\;B}\;.
\label{A2}	
\end{align} 
\end{widetext}
Here, $\Box$ denotes the d'Alembertian defined with respect to the standard covariant derivative, where the role 
of the affine connection is played by the Christoffel symbol. The terms including $\Box$ are proportional to a total divergence and 
can be discarded. The curvature tensor ${\cal R}_{\mu\nu\; A}^{\;\;\;B}$ is defined as follows:
\begin{align}
 	({\cal D}_{\mu}{\cal D}_{\nu}-
	 {\cal D}_{\nu}{\cal D}_{\mu})\psi_A =
	 {\cal R}_{\mu\nu\; A}^{\;\;\;B}\psi_B.
\label{curvature}
	\end{align}

Now we would like to fix the function $f^a(\varphi)$ for the gauge condition (\ref{GB1}). The gravity part of (\ref{GB1}), which coincides with the well-known de Donder gauge, implies the proportionality of the part being of second order in derivatives  
in the graviton-graviton block $\frac{\delta^2 S_{\rm tot}}
{\delta h_{\mu\nu} \delta h_{\alpha\beta}}$ of the operator $F_{AB}$ to the d'Alembertian \cite{Bar-Vil}. It is easy to see that the second variation of the action with respect to the scalar fields is also proportional to the d'Alembertian, independently of the form of the function $f^a(\varphi)$. The non-diagonal terms arise in the mixed functional derivatives and they have the form:
\begin{align}
 \frac{\delta S_{\rm tot}}{\delta h_{\alpha\beta} \delta \sigma^b} = 
U'\,n_b\,\nabla^{\alpha}\nabla^{\beta} + U\, f_b\, \nabla^{\alpha}\nabla^{\beta} + \cdots,
\label{GB2}
\end{align}
 where a ``prime'' denotes the derivative with respect to $\varphi$ and 
\begin{align}
 n_a \equiv \frac{\Phi_a}{\varphi}\;.
\label{n-def}
\end{align}
It follows immediately from (\ref{GB2})  that by choosing 
\begin{align}
 f_a = -\frac{U'}{U}n_a\;,
\label{GB3}
\end{align}
we obtain the proportionality of the operator $F_{AB}$ to the d'Alembertian. 

Now we write down the explicit expressions for the coefficients $\tilde{C}_{AB}, \Gamma_{AB}^{\mu}$ and $W_{AB}$ in Eq.
(\ref{second-order1}).\\
We have
\begin{align}
\tilde{C}_{AB}=\left( \begin{array}{ccc}
U\,G^{\alpha\beta\gamma\delta}&\quad &-\frac{1}{2}\,U'\,g^{\alpha\beta}\,n_{b}\\
\quad & \quad & \quad \\
-\frac{1}{2}\,U'\,n_{a}\,g^{\gamma\delta} &\quad &G\,\delta_{ab}+\frac{(U')^2}{U}\,n_{a}\,n_{b}\\\nonumber
\end{array}\right),\label{CSecDeriv}\\  
\end{align}
with the abbreviation $G_{\alpha\beta\gamma\delta}$ \cite{DeWitt} given by 
\begin{align}
G_{\alpha\beta\gamma\delta} \equiv& g_{\alpha\gamma}g_{\beta\delta}+g_{\alpha\delta}g_{\beta\gamma}-g_{\alpha\beta}g_{\gamma\delta}\;.
\label{supermetric1}
\end{align}
The inverse is
\begin{align}
G^{\alpha\beta\gamma\delta}=&\frac{1}{4}\Big( g^{\alpha\gamma}g^{\beta\delta}+g^{\alpha\delta}g^{\beta\gamma}-g^{\alpha\beta}g^{\gamma\delta}\Big)\;,\label{supermetric}
\end{align}
such that
\begin{align}
 G^{\alpha\beta\lambda\sigma}G_{\lambda\sigma\gamma\delta}=\delta^{\alpha\beta}_{\gamma\delta}\equiv\delta_{(\gamma}^{\alpha}\delta_{\delta)}^{\beta}=\frac{1}{2}\Big(\delta_{\gamma}^{\alpha}\delta_{\delta}^{\beta}+\delta_{\delta}^{\alpha}\delta_{\gamma}^{\beta}\Big)\;.
\end{align}
Then,
\begin{align}
\Gamma_{AB}^{\nu}=\left( \begin{array}{ccc}
\Gamma_{11}^{\nu}&\quad &\Gamma_{12}^{\nu}\\
\quad & \quad & \quad \\
 \Gamma_{21}^{\nu}&\quad &\Gamma_{22}^{\nu}\\
\end{array}\right),\label{Gamma0}\\\nonumber
\end{align}
where the subscript 1 stands for the graviton perturbation $h_{\mu\nu}$, while the subscript 2 denotes the scalar field perturbation
$\sigma^a$:
\begin{flalign}
\Gamma_{11}^{\nu}= &\,U'n_{a}\Phi^{a}_{;\,\mu}\Big(g^{\mu(\alpha}G^{\beta)\nu\gamma\delta}-g^{\mu(\gamma}G^{\delta)\nu\alpha\beta}+\frac{1}{2}\,g^{\nu\mu}G^{\alpha\beta\gamma\delta}\Big),\\
\Gamma_{12}^{\nu}= &\,G\,G^{\alpha\beta\mu\nu}\,\Phi_{a;\,\mu}+\frac{1}{2}\,U_{ab}\,\Phi^{b}_{;\,\mu}\Big(g^{\mu(\alpha}\,g^{\beta)\nu}-\frac{3}{2}\,g^{\alpha\beta}\,g^{\mu\nu}\Big),\\
\Gamma_{21}^{\nu}= &\,-G^{\gamma\delta\mu\nu}\,\Big(G\,\delta_{ab}+U_{ab}\Big)\,\Phi^{a}_{;\,\mu}\,,\quad\quad\quad\quad\quad\quad\quad\quad\\
\Gamma_{22}^{\nu}= &\,\frac{1}{2}\,G'\,g^{\mu\nu}\,\Phi^{c}_{;\,\mu}\,n^{d}\Big(-\delta_{bc}\delta_{ad}+\delta_{ac}\delta_{bd}+\delta_{ab}\delta_{cd}\Big)\nonumber\\
&+ \frac{1}{2}\,\frac{U'}{U}\Big(U_{bc}\,n_a+U_{ac}\,n_b-\frac{(U')^2}{U}\,n_a \,n_b \,n_c\Big)\, g^{\mu\nu}\,\Phi^{c}_{;\,\mu}\,.
\end{flalign}
Analogously, we have for $W_{AB}$:
\begin{flalign}
W_{11}= &\,U\,K^{\alpha\beta\gamma\delta}+G\,S^{\alpha\beta\gamma\delta}+VG^{\alpha\beta\gamma\delta}+T^{\alpha\beta\gamma\delta},\quad\quad\\
W_{12}= &\,G^{\alpha\beta\mu\nu}\Big(V'\,g_{\mu\nu}-2\,U'\,R_{\mu\nu}+G'\,\Phi^{c}_{,\,\mu}\,\Phi_{c,\,\nu}\Big)\,n_{b}\nonumber\\
&+\Big(g^{\mu(\alpha}g^{\beta)\nu}-g^{\alpha\beta}g^{\mu\nu}\Big)\Big(U_{bac}\,\Phi^{a}_{;\,\mu}\,\Phi^{c}_{;\,\nu}+U_{ba}\,\Phi^{a}_{;\,\nu\mu}\Big),\\
W_{21}= &\,n_{a}\,G^{\gamma\delta\mu\nu}\Big(V'\,g_{\mu\nu}-2\,U'\,R_{\mu\nu}+G'\,\Phi^{c}_{;\,\mu}\,\Phi_{c;\,\nu}\Big)\nonumber\\
&-2\,G^{\gamma\delta\mu\nu}\,G'\,n_{b}\,\Phi^{b}_{;\,\nu}\,\Phi_{a;\,\mu}-2\,G^{\gamma\delta\mu\nu}\,G\,\Phi_{a;\,\mu\nu}\;,\quad\\
W_{22}=            &\,U_{ab}\,R-\frac{1}{2}\,G_{ab}\,\Phi_{,\,\mu}^{c}\Phi^{,\,\mu}_{c}-V_{ab}+G_{cb}\,\Phi^{c}_{;\,\mu}\,\Phi^{;\,\mu}_{a}\nonumber\\
&+G'\,n_{b}(\Box\,\Phi_{a})\;.
\end{flalign}
with
\begin{flalign}
K^{\alpha\beta\gamma\delta} \equiv&\,-G^{\alpha\beta\gamma\delta}\,R-\frac{1}{2}g^{\alpha\beta}R^{\gamma\delta}-\frac{1}{2}g^{\gamma\delta}R^{\alpha\beta}+\frac{1}{2}g^{\alpha(\gamma}R^{\delta)\beta}\nonumber\\
&+\frac{1}{2}g^{\beta(\gamma}R^{\delta)\alpha}+\frac{1}{2}R^{\alpha\gamma\beta\delta}+\frac{1}{2}R^{\alpha\delta\beta\gamma}\;,\\\nonumber\\
S^{\alpha\beta\gamma\delta}\equiv &\,\frac{1}{2}G^{\alpha\beta\gamma\delta}\Phi^{a}_{;\,\mu}\Phi^{;\,\mu}_{a}+\frac{1}{4}g^{\alpha\beta}\Phi^{a;\,\gamma}\Phi_{a}^{;\,\delta}+\frac{1}{4}g^{\gamma\delta}\Phi^{a;\,\alpha}\Phi_{a}^{;\,\beta}\nonumber\\
&-\frac{1}{2}g^{\alpha(\gamma}\Phi^{a;\,\delta)}\Phi^{;\,\beta}_{a}-\frac{1}{2}g^{\beta(\gamma}\Phi^{a;\,\delta)}\Phi^{;\,\alpha}_{a}\;,\\\nonumber\\
T^{\alpha\beta\gamma\delta}\equiv &\,2\,G^{\alpha\beta\gamma\delta}\Box U+g^{\alpha\beta}U^{;\,\gamma\delta}+\frac{1}{2}g^{\gamma\delta}U^{;\,\alpha\beta}-g^{\alpha(\gamma}U^{;\,\delta)\beta}\nonumber\\
&-g^{\beta(\gamma}U^{;\,\delta)\alpha}\;.
\end{flalign}
The following formulae for a general function $Z(\varphi)$ will be useful:
\begin{flalign}
 \nabla_{\mu}\,Z= &\,Z'\,n_{a}\,\nabla_{\mu}\,\Phi^{a}\;,\\\nonumber\\
Z_{ab}\equiv& \,\frac{\partial^2 Z}{\partial \Phi^a\partial \Phi^b}= Z''\,n_a n_b+\frac{Z'}{\varphi}\Big(\delta_{ab}-n_a n_b\Big)\;,\\\nonumber\\
Z_{abc}\equiv &\,\frac{\partial^3\,Z}{\partial\Phi^a\partial\Phi^b\partial\Phi^c}=Z'''n_a n_b n_c+\frac{1}{\varphi}\Bigg[n_a\Big(Z''\delta_{bc}-Z_{bc}\Big)\nonumber\\
&+n_b\Big(Z''\delta_{ac}-Z_{ac}\Big)+n_c\Big(Z''\delta_{ab}-Z_{ab}\Big)\Bigg].
\end{flalign}
The next step in the algorithm described above is the calculation of the inverse matrix $\tilde{C}^{(-1)AB}$.
We obtain
\begin{widetext}
\begin{align}
 (\tilde{C}^{-1})^{BC}=\,\left( \begin{array}{ccc}
\frac{1}{U}\,G_{\lambda\sigma\gamma\delta}+\frac{(1-G\,s)}{3\,U}\,g_{\lambda\sigma}\,g_{\gamma\delta}&\quad &-s\,\frac{U'}{U}\,g_{\gamma\delta}\,n^{c}\\
\quad & \quad & \quad \\
-s\,\frac{U'}{U}\,n^{b}\,g_{\lambda\sigma} &\quad &\frac{1}{G}\Big(\delta^{bc}-n^{b}\,n^{c}\Big)+s\,n^{b}\,n^{c}\\
\end{array}\right),
\end{align} 
\end{widetext}
where the factor $s$ is given by the formula
\begin{align}
s \equiv\,\frac{U}{GU + 3U'^2}\;.
\label{s-define} 
\end{align}
It is exactly this factor that is responsible for the suppression of the contributions of quantum loop corrections, in particular for the suppression of the Higgs field 
propagators in models with a strong non-minimally coupled Higgs field \cite{Wil,BKKSS,BKKSS1,Clark,Lerner}.
\newpage
We abstain from writing down the results of the multiplication of the operator $F_{AB}$ by the inverse matrix  $\tilde{C}^{(-1)AB}$. By redefining the covariant derivative, we finally arrive at the minimal form (\ref{minim0}) of the operator $F_{A}^{\:B}$. 
Now we also have to find the Faddeev-Popov operator. Applying the definition (\ref{FP}), we obtain
\begin{align}
Q_{\alpha}^{\;\beta}=\,\delta^{\;\beta}_{\alpha}\,\Box+2\,\,{}^{^{(Q)}}\Gamma^{\:\beta\,\mu}_{ \alpha}\,\nabla_{\mu}+{}^{^{(Q)}}W_{\alpha}^{\;\beta}\;,
\label{FP1}
\end{align}
with
\begin{align}
{}^{^{(Q)}}\Gamma^{\;\beta\,\mu}_{\alpha}=\,-\frac{1}{2}\,\frac{U'}{U}\,n_{a}\,\Phi^{a}_{,\,\alpha}\,g^{\beta\mu}\;,
\end{align}
and
\begin{align}
{}^{^{(Q)}}W_{\alpha}^{\;\beta}=R^{\;\beta}_{\alpha}-\frac{U'}{U}\,n_{a}\,\nabla_{\alpha}\,\nabla^{\beta}\,\Phi^{a}.
\label{GhostPot}
\end{align} 
The Faddeev-Popov ghost operator (\ref{FP1}) also contains a term linear in the derivative. To absorb 
this term into the d'Alembertian we must again redefine the general affine connection for the ghost operator, just like it 
was done for the main operator $F_A^B$. As in the case of $F_A^B$ we refrain from writing down the explicit formula for the final minimal form of the Faddeev-Popov operator.\\ 
Now we are in a position to apply the formulae (\ref{div}) and (\ref{A2}). Since a vast number of terms are generated during the algorithm, we used (\ref{CSecDeriv})-(\ref{GhostPot}) as an input and have implemented the Schwinger-DeWitt algorithm in the MathTensor package \cite{Mathtensor} to calculate the final result for the divergent part of the one-loop effective action. This will be presented in the next section in a closed form. The result is rather lengthy, but we organize it in a ``physical'' way to facilitate its use. First of all, for practical purposes in a cosmological setup, the most important quantum corrections are $V_{1\text{-loop}}$, $U_{1\text{-loop}}$ and $G_{1\text{-loop}}$ and the result for those corrections is not very long. Then, we have arranged the result as a polynomial in the suppression function $s(\varphi)$, which is approximately zero for sufficiently high energies (as it is the case during inflation). However, the remaining structures, which are partly very long, are nevertheless very important from a physical perspective. As mentioned in the introduction, we adopt the viewpoint of an \emph{effective} field theory, which is valid up to a specific energy scale determined by a cutoff. The magnitude of this cutoff depends on the coefficients of the remaining structures, since they are supposed to be suppressed sufficiently by powers of this cutoff to guarantee the applicability of the model up to the corresponding energy scale.  
\onecolumngrid
\section{One-loop divergences in the Jordan frame: the result}

The divergent part of the one-loop effective action for a scalar multiplet non-minimally coupled to gravity in the Jordan frame can be written as the following sum of a minimal set of $21$ independent structures:
\begin{align}
 W^{\text{div}}_{\text{1-loop}}=&\frac{1}{32\,\pi^2\,(2-\omega)}\int\text{d}^4\,x\,\sqrt{g}\,\Big\{\alpha_{1}+\alpha_{2}\,R+\frac{1}{2}\,\alpha_{3}\,\Phi^{a}_{,\,\mu}\Phi_{a}^{,\,\mu}+\alpha_{4}\,\Phi^{a}_{,\,\mu}n_{a}\Phi^{b\,,\,\mu}n_{b}+\alpha_{5}\,R^{\mu\nu}R_{\mu\nu}+\alpha_{6}\,R^2\nonumber\\\nonumber\\\nonumber
&\quad\quad\quad\quad\quad\quad+\alpha_{7}\,R^{\mu\nu}\Phi^{a}_{,\,\mu}\Phi_{a,\,\nu}+\alpha_{8}\,R^{\mu\nu}\Phi^{a}_{,\,\mu}n_{a}\Phi^{b}_{,\,\nu}n_{b}+\alpha_{9}\,R\,(\Phi^{a}_{,\,\mu}\Phi_{a}^{,\,\mu})+\alpha_{10}\,R\,(\Phi^{a}_{,\,\mu}n_{a}\Phi^{b\,,\,\mu}n_{b})+\alpha_{11}\,R\,(\Phi^{c\;\;\mu}_{;\,\mu}\,n_{c})\\\nonumber\\\nonumber
&\quad\quad\quad\quad\quad\quad+\alpha_{12}\,(\Phi^{a}_{,\,\mu}\Phi_{a}^{,\,\mu})^2+\alpha_{13}\,(\Phi^{a}_{,\,\mu}n_{a}\Phi^{b\,,\,\mu}n_{b})^2+\alpha_{14}\,(\Phi^{a}_{,\,\mu}\Phi_{a}^{,\,\mu})\,(\Phi^{c}_{,\,\nu}n_{c}\Phi^{d\,,\,\nu}n_{d})+\alpha_{15}\,\Phi^{a}_{,\,\mu}\Phi_{a\,,\,\nu}\Phi^{b\,,\,\mu}\Phi_{b}^{,\,\nu}\\\nonumber\\\nonumber
&\quad\quad\quad\quad\quad\quad+\alpha_{16}\,\Phi^{a}_{,\mu}n_{a}\Phi^{b}_{,\nu}n_{b}\Phi^{c\;,\,\mu}\Phi_{c}^{,\,\nu}+\alpha_{17}\,\Phi^{a\;\;\mu}_{;\,\mu}\Phi_{a;\,\nu}^{\;\;\;\;\;\nu}+\alpha_{18}\,(\Phi^{c\;\;\mu}_{;\,\mu}\,n_{c})^2+\alpha_{19}(\Phi^{a}_{,\,\mu}\Phi_{a}^{,\,\mu})(\Phi^{b\;\;\nu}_{;\,\nu}n_{b})\\\nonumber\\
&\quad\quad\quad\quad\quad\quad+\alpha_{20}\,(\Phi^{a}_{,\,\mu}n_{a}\Phi^{b\,,\,\mu}n_{b})(\Phi^{c\;\;\nu}_{;\,\nu}\,n_{c})+\alpha_{21}\,\Phi^{a\;\;\mu}_{;\,\mu}\Phi_{a}^{,\,\nu}\Phi^{b}_{,\,\nu}n_{b}\Big\}\;.\label{effectiveaction}
\end{align}
The structures in (\ref{effectiveaction}) are ordered with respect to an increasing number of derivatives and furthermore sorted by pure curvature terms, mixed (curvature and gradient) terms and pure gradient terms. We have taken into account integration by parts and the Gau\ss-Bonnet identity to eleminate $8$ dependent structures and to arrive at the minimal set of $21$ independent structures (for the explicit formulae, see (\ref{transferequation1})-(\ref{transfer equations}) in Appendix A).

To collect the contributions in each coefficient $\alpha_{i}$ in a systematical way, we have adopted the following sorting pattern:
First of all, the contributions are ordered in terms of powers of the potential $V$ and its derivatives $V',V'',V'''$. Then, we have further rewritten the coefficients as a polynomial in decreasing powers of $s(\varphi)$. Expressed in this form one can deduce direct physical information for cosmological applications as e.g.~in the context of the Higgs inflation scenario. The tree level values of $U,\,G,\,V$ in this model have the concrete form of eq. (\ref{treelevelchargeU}),~(\ref{treelevelchargeG}),~(\ref{treelevelchargeV}). For inflationary energy scales, corresponding to field values $\varphi\gg M_{\text{P}}/\sqrt{\xi}$, 
 the ``suppression function'' $s$ scales as $s=1/6\xi$. Since the non-minimal coupling constant $\xi$ is of order $10^4-10^5$ in this kind of models, it follows $s\simeq0$. Thus, in the inflationary context of a scalar multiplet non-minimally coupled to gravity most of the terms in the coefficients $\alpha_{i}$ are sufficiently suppressed and can be neglected.

All structures $\alpha_{i}$ which contain more than two derivatives are supposed to be suppressed by the appropriate power of some cutoff $\Lambda$, which determines the validity of the model. All such structures appearing in (\ref{effectiveaction}) can be denoted  symbolically as
\begin{align}
 \frac{RR}{\Lambda^2},\quad\quad\frac{R\,\partial^2\Phi\Phi}{\Lambda^2},\quad\quad\frac{\partial^4\Phi\Phi\Phi\Phi}{\Lambda^2}\;,
\end{align}
corresponding to pure curvature structures, mixed structures or pure scalar gradient structures. The structures $\alpha_{i},\,i=12,...,21$ only contain four derivatives of the scalar fields and belong all to the pure gradient structures. In \cite{BKKSS1} it was shown that those structures are suppressed by an additional small factor, the slow roll parameter  $\hat{\epsilon}\ll 1$, compared to the curvature structures, i.e.
\begin{align}
 \frac{R}{\Lambda^2}\simeq\frac{\lambda}{16\pi^2}\quad\text{compared to}\quad\frac{\partial}{\Lambda}\simeq\frac{\sqrt{\lambda}}{48\pi}\sqrt{2\hat{\epsilon}}\;,
\end{align}
with the Higgs self-coupling $\lambda\simeq10^{-1}$. Thus, the suppression mechanism for the gradient structures works even more efficiently. From an effective field theory point of view, these terms are less important and therefore shifted to the Appendix B. Below, we present only the result for the coefficients $\alpha_{i},\,i=1,...,11$. The result is given in a closed form, beginning with the most important structures $\alpha_{1},\,\alpha_{2}$ and $\alpha_{3}$ corresponding to $V_{1\text{-loop}}$, $U_{1\text{-loop}}$ and $G_{1\text{-loop}}$. The factor of $1/2$ in front of $G_{1\text{-loop}}$ is due to the form of the kinetic term $\frac{1}{2}\Phi^{a}_{;\,\mu}\Phi_{a}^{;\,\mu}$ in the action (\ref{action}). \\\\\\ 

The one-loop correction to the potential $V(\varphi)$ :

\begin{align}
\alpha_{1}=&\;V^2
   \left[\frac{2 s^2 \left(U'\right)^4}{U^4}-\frac{2 s
   \left(U'\right)^2}{U^3}+\frac{5}{U^2}\right]+V V'
   \left[-\frac{8 s^2
   \left(U'\right)^3}{U^3}+\frac{4 s U'}{U^2}\right]+V\,V''\,\frac{2  s^2
   \left(U'\right)^2 }{U^2}\nonumber\\
&+\left(V'\right)^2 \left[\frac{8 s^2
   \left(U'\right)^2}{U^2}-\frac{2 s}{U}+\frac{(N-1)}{2 G^2 \varphi ^2}\right]- V' V''\,\frac{4 s^2 U'}{U}+\frac{1}{2} \left(V''\right)^2 s^2
   \label{V1loop}
\end{align}
\\

The one-loop correction to the non-minimal coupling $U(\varphi)$ :

\begin{align}
\alpha_{2}=&\;V
   \left[s^2 \left(\frac{4 \left(U'\right)^4}{U^3}-\frac{2
   \left(U'\right)^2 U''}{U^2}\right)-\frac{7 s
   \left(U'\right)^2}{3 U^2}-\frac{13}{3 U}\right]+V' \left[s^2 \left(\frac{4 U' U''}{U}-\frac{8
   \left(U'\right)^3}{U^2}\right)+\frac{8 s U'}{3 U}-\frac{(N-1) U'}{G^2 \varphi ^2}-\frac{(N-1)}{6 G \varphi
   }\right]\nonumber\\
&+V'' \left[s^2
   \left(\frac{2
   \left(U'\right)^2}{U}-U''\right)-\frac{s}{6}\right]
    \label{U1loop}
\end{align}
\\

The one-loop correction to the kinetic term $\frac{1}{2}\,G(\varphi)$ :
\begin{align}
 \frac{1}{2}\,\alpha_{3}=&\; V \Bigg[s^2
   \Bigg(\frac{\left(U'\right)^6}{2 G U^4 \varphi ^2}+\frac{2 G'
   \left(U'\right)^2}{U^2 \varphi }+\frac{G''
   \left(U'\right)^2}{U^2}-\frac{G'
   \left(U'\right)^3}{U^3}-\frac{\left(G'\right)^2
   \left(U'\right)^2}{2 G U^2}+\frac{6 \left(U'\right)^5}{U^4
   \varphi }+\frac{\left(U'\right)^4}{6 U^3 \varphi ^2}+\frac{3
   \left(U'\right)^6}{2 U^5}\Bigg)\nonumber\\
&\quad\quad+s
   \left(-\frac{\left(U'\right)^4}{3 G U^3 \varphi ^2}-\frac{G'
   \left(U'\right)^2}{2 G U^2 \varphi }+\frac{G' U'}{2
   U^2}-\frac{9 \left(U'\right)^3}{2 U^3 \varphi }-\frac{19
   \left(U'\right)^2}{18 U^2 \varphi ^2}-\frac{2
   \left(U'\right)^4}{U^4}\right)+\frac{19 \left(U'\right)^2}{18
   G U^2 \varphi ^2}+\frac{G}{U^2}+\frac{3 U'}{U^2 \varphi
   }+\frac{\left(U'\right)^2}{2 U^3}\Bigg]\nonumber\\
&+V' \Bigg[s^2 \Bigg(-\frac{3
   \left(U'\right)^5}{G U^3 \varphi ^2}-\frac{3 G'
   \left(U'\right)^3}{G U^2 \varphi }-\frac{5 G' U'}{U \varphi
   }-\frac{2 G'' U'}{U}+\frac{2 G'
   \left(U'\right)^2}{U^2}+\frac{\left(G'\right)^2 U'}{G
   U}-\frac{12 \left(U'\right)^4}{U^3 \varphi
   }-\frac{\left(U'\right)^3}{U^2 \varphi ^2}-\frac{3
   \left(U'\right)^5}{U^4}\Bigg)\nonumber\\
&\quad\quad+s \Bigg(\frac{5
   \left(U'\right)^4}{4 G^2 U^2 \varphi
   ^3}+\frac{\left(U'\right)^3}{2 G U^2 \varphi ^2}+\frac{3 G'
   \left(U'\right)^2}{2 G^2 U \varphi ^2}+\frac{41
   \left(U'\right)^2}{12 G U \varphi ^3}+\frac{2 G' U'}{G U
   \varphi }-\frac{G'}{2 U}-\frac{\left(G'\right)^2}{4 G^2
   \varphi }-\frac{G'}{2 G \varphi ^2}+\frac{25
   \left(U'\right)^2}{4 U^2 \varphi }+\frac{11 U'}{6 U \varphi
   ^2}\nonumber\\
&\quad\quad\quad+\frac{5 \left(U'\right)^3}{2 U^3}\Bigg)-\frac{5
   \left(U'\right)^2}{12 G^2 U \varphi ^3}-\frac{11 U'}{6 G U
   \varphi ^2}-\frac{G'}{2 G^2 \varphi ^2}-\frac{2}{U \varphi
   }-\frac{U'}{2 U^2}+\frac{(N-1) G'}{2 G^2 \varphi ^2}\Bigg]\nonumber\\
&+V'' \Bigg[s^2
   \left(\frac{3 G' \left(U'\right)^2}{2 G U \varphi }-\frac{G'
   U'}{2 U}+\frac{3 G'}{2 \varphi }-\frac{\left(G'\right)^2}{4
   G}+\frac{G''}{2}+\frac{3 \left(U'\right)^3}{U^2 \varphi
   }+\frac{3 \left(U'\right)^4}{4 U^3}\right)+s
   \left(-\frac{G'}{2 G \varphi }-\frac{U'}{U \varphi
   }-\frac{\left(U'\right)^2}{4 U^2}\right)\Bigg]
    \label{G1loop}
\end{align}
\\

The coefficient $ \alpha_{4}$ in front of the structure $\Phi^{a}_{,\,\mu}n_{a}\Phi^{b\,,\,\mu}n_{b}$ :
\begin{align}
 \alpha_{4}=&\; V \;\Bigg[\; s^3\left(\frac{99 \left(U'\right)^8}{2
   U^6}-\frac{90 U'' \left(U'\right)^6}{U^5}-\frac{9 G'
   \left(U'\right)^5}{U^4}-\frac{18 \left(U''\right)^2
   \left(U'\right)^4}{U^4}-\frac{18 G' U''
   \left(U'\right)^3}{U^3}-\frac{5 \left(G'\right)^2
   \left(U'\right)^2}{2 U^2}\right)\nonumber\\
  &\quad+ s^2\Bigg(-\frac{\left(U'\right)^6}{2 G \varphi ^2
   U^4}-\frac{54 \left(U'\right)^6}{U^5}-\frac{6
   \left(U'\right)^5}{\varphi  U^4}+\frac{147 U''
   \left(U'\right)^4}{2 U^4}-\frac{\left(U'\right)^4}{6 \varphi
   ^2 U^3}+\frac{13 G' \left(U'\right)^3}{2
   U^3}+\frac{\left(G'\right)^2 \left(U'\right)^2}{2 G
   U^2}+\frac{9 \left(U''\right)^2
   \left(U'\right)^2}{U^3}\nonumber\\
&\quad\quad\quad-\frac{2 G' \left(U'\right)^2}{\varphi
    U^2}+\frac{5 G' U'' U'}{2 U^2}\Bigg)+s\left(\frac{\left(U'\right)^4}{3 G \varphi ^2
   U^3}+\frac{41 \left(U'\right)^4}{2 U^4}+\frac{9
   \left(U'\right)^3}{2 \varphi  U^3}+\frac{G'
   \left(U'\right)^2}{2 G \varphi  U^2}-\frac{17 U''
   \left(U'\right)^2}{U^3}+\frac{19 \left(U'\right)^2}{18
   \varphi ^2 U^2}-\frac{G' U'}{U^2}\right)\nonumber\\
&\quad -\frac{19
   \left(U'\right)^2}{18 G \varphi ^2 U^2}+\frac{33
   \left(U'\right)^2}{2 U^3}-\frac{3 U'}{\varphi  U^2}+\frac{3
   U''}{U^2}\Bigg]\nonumber\\
&+V' \Bigg[s^3\Bigg(-\frac{99 \left(U'\right)^7}{U^5}+\frac{180 U''
   \left(U'\right)^5}{U^4}+\frac{18 G'
   \left(U'\right)^4}{U^3}+\frac{36 \left(U''\right)^2
   \left(U'\right)^3}{U^3}+\frac{36 G' U''
   \left(U'\right)^2}{U^2}+\frac{5 \left(G'\right)^2
   U'}{U}\Bigg)+s^2\Bigg(\frac{3 \left(U'\right)^5}{G \varphi
   ^2 U^3}\nonumber\\
&\quad+\frac{165 \left(U'\right)^5}{2 U^4}+\frac{12
   \left(U'\right)^4}{\varphi  U^3}+\frac{3 G'
   \left(U'\right)^3}{G \varphi  U^2}-\frac{195 U''
   \left(U'\right)^3}{2 U^3}+\frac{\left(U'\right)^3}{\varphi ^2
   U^2}\nonumber-\frac{17 G' \left(U'\right)^2}{2
   U^2}-\frac{\left(G'\right)^2 U'}{G U}-\frac{15
   \left(U''\right)^2 U'}{U^2}+\frac{5 G' U'}{\varphi 
   U}\nonumber\\
&\quad-\frac{5 G' U''}{2 U}\Bigg)+s\Bigg(-\frac{5
   \left(U'\right)^4}{4 G^2 \varphi ^3
   U^2}-\frac{\left(U'\right)^3}{2 G \varphi ^2 U^2}-\frac{26
   \left(U'\right)^3}{U^3}-\frac{3 G' \left(U'\right)^2}{2 G^2
   \varphi ^2 U}-\frac{41 \left(U'\right)^2}{12 G \varphi ^3
   U}-\frac{25 \left(U'\right)^2}{4 \varphi  U^2}-\frac{2 G'
   U'}{G \varphi  U}+\frac{14 U'' U'}{U^2}\nonumber\\
&\quad-\frac{11 U'}{6
   \varphi ^2 U}+\frac{\left(G'\right)^2}{4 G^2 \varphi
   }+\frac{G'}{U}+\frac{G'}{2 G \varphi ^2}\Bigg)+\frac{5
   \left(U'\right)^2}{12 G^2 \varphi ^3 U}+\frac{G'}{2 G^2 \varphi ^2}+\frac{11 U'}{6 G
   \varphi ^2 U}-\frac{5 U'}{2 U^2}+\frac{2}{\varphi 
   U}+\frac{(N-1) G'}{2 G^2
   \varphi ^2}+\frac{3
   (N-1) \left(G'\right)^2}{4 G^3 \varphi }\Bigg]\nonumber\\
&\;+ V''\Bigg[s^3\Bigg(\frac{99 \left(U'\right)^6}{4
   U^4}-\frac{45 U'' \left(U'\right)^4}{U^3}-\frac{9 G'
   \left(U'\right)^3}{2 U^2}-\frac{9 \left(U''\right)^2
   \left(U'\right)^2}{U^2}-\frac{9 G' U'' U'}{U}-\frac{5
   \left(G'\right)^2}{4}\Bigg)+ s^2\Bigg(-\frac{111
   \left(U'\right)^4}{4 U^3}-\frac{3 \left(U'\right)^3}{\varphi 
   U^2}\nonumber\\
&\quad-\frac{3 G' \left(U'\right)^2}{2 G \varphi  U}+\frac{12
   U'' \left(U'\right)^2}{U^2}-\frac{G' U'}{2
   U}+\frac{\left(G'\right)^2}{4 G}+\frac{3
   \left(U''\right)^2}{U}-\frac{3 G'}{2 \varphi }\Bigg)+s\left(\frac{45 \left(U'\right)^2}{4
   U^2}+\frac{U'}{\varphi  U}+\frac{G'}{2 G \varphi
   }-\frac{U''}{U}\right) -\frac{2}{U}\nonumber\\
&\quad-\frac{(N-1) G'}{2 G^2 \varphi
   }\Bigg]+V'''\Bigg[s^2 \left(\frac{9
   \left(U'\right)^3}{2 U^2}+\frac{G'}{2}\right)-\frac{s
   U'}{U}\Bigg] 
\end{align}
\\

The coefficient $ \alpha_{5}$ in front of the structure $R^{\mu\nu}R_{\mu\nu}$ :
\begin{align}
 \alpha_{5}=&\;\frac{2 s \left(U'\right)^2}{U}+\frac{43}{60}
\end{align}
\\

The coefficient $ \alpha_{6}$ in front of the structure $R^2$ :
\begin{align}
 \alpha_{6}=&\; s^2 \left(-\frac{2
   \left(U'\right)^2 U''}{U}+\frac{2
   \left(U'\right)^4}{U^2}+\frac{\left(U''\right)^2}{2}\right)+s
   \left(\frac{U''}{6}-\frac{4 \left(U'\right)^2}{3
   U}\right)+\frac{1}{40}+\frac{(N-1)}{72}+\frac{(N-1)
   U'}{6 G \varphi }+\frac{(N-1) \left(U'\right)^2}{2 G^2 \varphi ^2}
\end{align}
\\

The coefficient $ \alpha_{7}$ in front of the structure $R^{\mu\nu}\Phi^{a}_{,\,\mu}\Phi_{a,\,\nu}$ :
\begin{align}
 \alpha_{7}=&\; s \left(-\frac{3 \left(U'\right)^4}{G U^2 \varphi ^2}+\frac{G'
   \left(U'\right)^2}{G U \varphi }-\frac{G' U'}{U}+\frac{5
   \left(U'\right)^3}{U^2 \varphi }+\frac{\left(U'\right)^2}{U
   \varphi ^2}+\frac{3
   \left(U'\right)^4}{U^3}\right)+\frac{\left(U'\right)^2}{G U
   \varphi ^2}-\frac{2 U'}{U \varphi
   }-\frac{\left(U'\right)^2}{U^2}
\end{align}
\\

The coefficient $ \alpha_{8}$ in front of the structure $R^{\mu\nu}\Phi^{a}_{,\,\mu}n_{a}\Phi^{b}_{,\,\nu}n_{b}$ :
\begin{align}
 \alpha_{8}=&\; s^2 \left(-\frac{G' U'
   U''}{U}-\frac{3 G' \left(U'\right)^3}{U^2}-\frac{15
   \left(U'\right)^4 U''}{U^3}-\frac{6 \left(U'\right)^2
   \left(U''\right)^2}{U^2}+\frac{9
   \left(U'\right)^6}{U^4}\right)+ s \Bigg(\frac{3 \left(U'\right)^4}{G U^2 \varphi ^2}-\frac{G'
   \left(U'\right)^2}{G U \varphi }\nonumber\\
&\;+\frac{2 G' U'}{U}-\frac{5
   \left(U'\right)^3}{U^2 \varphi }-\frac{\left(U'\right)^2}{U
   \varphi ^2}+\frac{10 \left(U'\right)^2 U''}{U^2}-\frac{8
   \left(U'\right)^4}{U^3}+\frac{2
   \left(U''\right)^2}{U}\Bigg)-\frac{\left(U'\right)^2}{G U
   \varphi ^2}+\frac{2 U'}{U \varphi }-\frac{2 U''}{U}
\end{align}

The coefficient $ \alpha_{9}$ in front of the structure $R\,(\Phi^{a}_{,\,\mu}\Phi_{a}^{,\,\mu})$ :
\begin{align}
 \alpha_{9}=&\; s^2 \Bigg(-\frac{3 G'
   \left(U'\right)^2 U''}{2 G U \varphi }-\frac{G' U''}{\varphi
   }+\frac{3 \left(U'\right)^6}{2 G U^3 \varphi ^2}+\frac{3 G'
   \left(U'\right)^4}{2 G U^2 \varphi }+\frac{3 G'
   \left(U'\right)^2}{2 U \varphi }+\frac{G''
   \left(U'\right)^2}{U}+\frac{G' U' U''}{2
   U}+\frac{\left(G'\right)^2 U''}{4 G}-\frac{G'
   \left(U'\right)^3}{U^2}\nonumber\\
&\quad-\frac{\left(G'\right)^2
   \left(U'\right)^2}{2 G U}-\frac{G'' U''}{2}+\frac{3
   \left(U'\right)^3 U''}{2 U^2 \varphi }-\frac{3
   \left(U'\right)^5}{U^3 \varphi }+\frac{\left(U'\right)^4}{2
   U^2 \varphi ^2}-\frac{3 \left(U'\right)^4 U''}{4 U^3}+\frac{3
   \left(U'\right)^6}{2 U^4}\Bigg)+s \Bigg(\frac{G' U''}{2 G \varphi }-\frac{5
   \left(U'\right)^5}{4 G^2 U^2 \varphi
   ^3}\nonumber\\
&\quad+\frac{\left(U'\right)^4}{12 G U^2 \varphi ^2}-\frac{3 G'
   \left(U'\right)^3}{2 G^2 U \varphi ^2}-\frac{41
   \left(U'\right)^3}{12 G U \varphi ^3}-\frac{2 G'
   \left(U'\right)^2}{G U \varphi }+\frac{\left(G'\right)^2
   U'}{4 G^2 \varphi }+\frac{G' U'}{2 G \varphi ^2}+\frac{G'
   U'}{2 U}-\frac{G'}{12 \varphi }+\frac{\left(G'\right)^2}{12
   G}-\frac{G''}{12}+\frac{\left(U'\right)^3}{2 U^2 \varphi
   }\nonumber\\
&\quad-\frac{83 \left(U'\right)^2}{36 U \varphi
   ^2}+\frac{\left(U'\right)^2 U''}{4 U^2}-\frac{9
   \left(U'\right)^4}{4 U^3}\Bigg)+\frac{5
   \left(U'\right)^3}{12 G^2 U \varphi ^3}+\frac{47
   \left(U'\right)^2}{36 G U \varphi ^2}+\frac{G' U'}{2 G^2
   \varphi ^2}-\frac{G}{3 U}+\frac{G'}{6 G \varphi }-\frac{2
   U'}{U \varphi }+\frac{7 \left(U'\right)^2}{12 U^2}\nonumber\\
&\quad-\frac{(N-1) G' U'}{G^2 \varphi ^2}-\frac{(N-1) G'}{6 G \varphi
   }
\end{align}
\\

The coefficient $ \alpha_{10}$ in front of the structure $R\,(\Phi^{a}_{,\,\mu}n_{a}\Phi^{b\,,\,\mu}n_{b})$ :
\begin{align}
 \alpha_{10}=&\; s^3\Bigg(-\frac{9 \left(U'\right)^8}{2 U^5}+\frac{81 U''
   \left(U'\right)^6}{4 U^4}+\frac{3 G'
   \left(U'\right)^5}{U^3}-\frac{27 \left(U''\right)^2
   \left(U'\right)^4}{U^3}-\frac{15 G' U'' \left(U'\right)^3}{2
   U^2}+\frac{9 \left(U''\right)^3
   \left(U'\right)^2}{U^2}-\frac{\left(G'\right)^2
   \left(U'\right)^2}{2 U}\nonumber\\
&\quad+\frac{3 G' \left(U''\right)^2
   U'}{U}+\frac{1}{4} \left(G'\right)^2 U''\Bigg)+s^2\Bigg(-\frac{3 \left(U'\right)^6}{2 G \varphi ^2
   U^3}+\frac{3 \left(U'\right)^6}{8 U^4}+\frac{3
   \left(U'\right)^5}{\varphi  U^3}-\frac{3 G'
   \left(U'\right)^4}{2 G \varphi  U^2}-\frac{45 U''
   \left(U'\right)^4}{4 U^3}-\frac{\left(U'\right)^4}{2 \varphi
   ^2 U^2}\nonumber\\
&\quad+\frac{9 G' \left(U'\right)^3}{4 U^2}-\frac{3 U''
   \left(U'\right)^3}{2 \varphi  U^2}+\frac{\left(G'\right)^2
   \left(U'\right)^2}{2 G U}+\frac{18 \left(U''\right)^2
   \left(U'\right)^2}{U^2}-\frac{3 G' \left(U'\right)^2}{2
   \varphi  U}-\frac{G'' \left(U'\right)^2}{U}+\frac{3 G' U''
   \left(U'\right)^2}{2 G \varphi  U}+\frac{G' U'' U'}{2
   U}\nonumber\\
&\quad-\frac{3
   \left(U''\right)^3}{U}+\frac{\left(G'\right)^2}{24}-\frac{\left(G'\right)^2 U''}{4 G}+\frac{G' U''}{\varphi }+\frac{G''
   U''}{2}\Bigg) +s\Bigg(\frac{5 \left(U'\right)^5}{4 G^2
   \varphi ^3 U^2}-\frac{\left(U'\right)^4}{12 G \varphi ^2
   U^2}+\frac{11 \left(U'\right)^4}{4 U^3}+\frac{3 G'
   \left(U'\right)^3}{2 G^2 \varphi ^2 U}\nonumber\\
&\quad+\frac{41
   \left(U'\right)^3}{12 G \varphi ^3
   U}-\frac{\left(U'\right)^3}{2 \varphi  U^2}+\frac{2 G'
   \left(U'\right)^2}{G \varphi  U}-\frac{2 U''
   \left(U'\right)^2}{U^2}+\frac{83 \left(U'\right)^2}{36
   \varphi ^2 U}-\frac{\left(G'\right)^2 U'}{4 G^2 \varphi
   }-\frac{G' U'}{U}-\frac{G' U'}{2 G \varphi
   ^2}-\frac{\left(G'\right)^2}{12
   G}-\frac{\left(U''\right)^2}{U}\nonumber\\
&\quad+\frac{G'}{12 \varphi
   }+\frac{G''}{12}-\frac{G' U''}{2 G \varphi }\Bigg)-\frac{5
   \left(U'\right)^3}{12 G^2 \varphi ^3 U}-\frac{47 \left(U'\right)^2}{36 G
   \varphi ^2 U}-\frac{16 \left(U'\right)^2}{3 U^2}-\frac{G'}{6 G \varphi }-\frac{G' U'}{2 G^2 \varphi ^2}+\frac{2
   U'}{\varphi  U}-\frac{2 U''}{U}\nonumber\\
&\quad+\frac{(N-1)
   G'}{12 G \varphi }+\frac{(N-1) G' U'}{2
   G^2 \varphi ^2}+\frac{(N-1)
   \left(G'\right)^2}{24 G^2}+\frac{(N-1)
   \left(G'\right)^2 U'}{4 G^3 \varphi }-\frac{(N-1) G''}{12 G}-\frac{(N-1) U' G''}{2
   G^2 \varphi }
\end{align}
\\

The coefficient $ \alpha_{11}$ in front of the structure $R\,(\Phi^{c\;\;\mu}_{;\,\mu}\,n_{c})$ :
\begin{align}
 \alpha_{11}=&\; s^2
   \left(\frac{G' U''}{2}-\frac{G' \left(U'\right)^2}{U}+\frac{9
   \left(U'\right)^3 U''}{2 U^2}-\frac{9
   \left(U'\right)^5}{U^3}\right)+s \left(\frac{G'}{12}-\frac{U'
   U''}{U}+\frac{19 \left(U'\right)^3}{4 U^2}\right)-\frac{3
   U'}{U}\nonumber\\
&\quad-\frac{(N-1) G'}{12 G}-\frac{(N-1) G' U'}{2 G^2 \varphi }\label{alphaeleven}
\end{align}
\\
In this section we have presented the ``off-shell" result for the most important terms of the divergent part of the one-loop effective action in a closed form. The remaining structures can be found in Appendix B.

 We did not use the equations of motion yet. This will be done in an upcoming publication \cite{we-future}. There, it will be of interest whether the results for the effective action calculated in the Jordan and Einstein frame coincide at least ``on-shell''. Following \cite{Tyutin} and also the unique effective action approach developed in \cite{Vilkovisky}, this should be the case.

We want to emphazise that the result presented here in its full length will serve as a starting point for further studies. The main consequences which follow from this result will be discussed in the upcoming publications \cite{we-future, we-future1}. However, one can already get some physical information from the result presented here, since all terms are arranged with an eye towards cosmological applications. Therefore, the top priority was given to the most important one-loop structures $V,U$ and $G$ and, furthermore, the coefficients are ordered according to powers of the potential $V$ and its derivatives and powers of the suppression function $s(\varphi)$ to faciliate the use of the result for cosmological applications.\vspace{1cm}\\

\twocolumngrid

\section{Checks and Comparison with Known Results}
Since our result is rather cumbersome it makes sense to apply all kinds of consistency checks and tests that are possible. One can distinguish between internal consistency tests such as the correct dimensionality and the scaling behaviour of certain quantities and cross-checks by comparing our general result for some limiting cases with results present in the literature. Fortunately, these checks can isolate certain features of our general model and thus test complementary aspects such as the consequences due to the non-minimal coupling, or the influence due to the presence of the $O(N)$ multiplet.

\subsection{Test of Dimensional and Scaling Behaviour}

\subsubsection{Dimensional Tests}
Writing the mass dimension of some object $O$ as $[O]_{M}$, one obvious test of the 1-loop results is that for all terms 
\begin{align}
[\alpha_{i}\cdot \Gamma_{i}]_{M}=4
\end{align}
should hold ($c=\hbar=1$). We have introduced the symbol $\Gamma_{i}$ for the structure corresponding to the coefficient $\alpha_{i}$, i.e.
\begin{align}
\Gamma_{1\text{-loop}}^{\text{div}}=\frac{1}{32\,\pi^2\,(2-\omega)}\int\text{d}^4x\,\sqrt{g}\,\sum_{i}\alpha_{i}\,\Gamma_{i}\;.
\end{align}
For this check it is useful to take the convention that $[\varphi]_{M}=1$, $[g_{\mu\nu}]_{M}=[g^{\mu\nu}]_{M}=0$. Note that we have chosen to shift the dimensionality in the coordinates instead of the metric. This means that each space-time derivative $[\partial_{\mu}]_{M}=1$ increases the mass dimension by one. For this convention it follows that $[U(\varphi)]_{M}=2$, $[G(\varphi)]_{M}=0$ and $[V(\varphi)]_{M}=4$. Each derivative of $U,V$ and $G$ with respect to $\varphi$ reduces the mass dimension by one and therefore $[s(\varphi)]_{M}=0$. This check was done successfully for all structures.

\subsubsection{Scaling Behaviour}
Another test is the scaling behaviour of the $\alpha_{i}$. The coefficients $\alpha_{i}$ are functions of $U,G,V$ and their derivatives $U^{(n)},G^{(n)},V^{(n)}$, 
\begin{align}
 \alpha_{i}(U^{(n)},G^{(n)},V^{(n)}),\;n=0,...\,,k\;.
\end{align}
The structure of the Schwinger-DeWitt algorithm forces them to be homogeneous functions of degree zero in their arguments, i.e.~they should behave as
\begin{align}
 \alpha_{i}(\lambda U^{(n)},\lambda G^{(n)},\lambda V^{(n)})= \alpha_{i}( U^{(n)}, G^{(n)},V^{(n)})
\end{align}
under the scaling
\begin{align}
 U^{(n)},G^{(n)},V^{(n)}\rightarrow \lambda U^{(n)},\lambda G^{(n)},\lambda V^{(n)}
\end{align}
for some constant $\lambda$.
We have explicitly checked that all $\alpha_{i}$ share this property.

\subsubsection{$O(N)$ Symmetry and Single Field Limit}
A third test can be made due to the $O(N)$ symmetry. The space-time derivatives of the normal vector $n_{a}$ generate powers of $\varphi$ in the denominator of the $\alpha_{i}$. In addition we have structures which are proportional to $(N-1)$ due to the projector $(\delta_{ab}-n_{a}n_{b})$.\\
In the single field limit we have $\Phi^{a}\rightarrow\varphi$ and $\delta_{ab}=n_{a}=N\rightarrow1$. This leads to a degeneracy of certain structures $\Gamma_{i}$, listed in the table below, which collapse into the same structures appearing in the effective action of \cite{shapiro}.
\begin{table}[h]
\begin{center}
\begin{tabular}{|l|l|}\hline
$O(N)$\text{ multiplet}&\text{single field}\\\hline
$\Gamma_{1}$&$\Gamma_{1}$\\\hline
$\Gamma_{2}$&$\Gamma_{2}$\\\hline
$\Gamma_{3},\Gamma_{4}$&$\Gamma_{3}$\\\hline
$\Gamma_{5}$&$\Gamma_{5}$\\\hline
$\Gamma_{6}$&$\Gamma_{6}$\\\hline
$\Gamma_{7},\Gamma_{8}$&$\Gamma_{7}$\\\hline
$\Gamma_{9},\Gamma_{10}$&$\Gamma_{9}$\\\hline
$\Gamma_{11}$&$\Gamma_{11}$\\\hline
$\Gamma_{12},\Gamma_{13},\Gamma_{14},\Gamma_{15}, \Gamma_{16}$&$\Gamma_{12}$\\\hline
$\Gamma_{17},\Gamma_{18}$&$\Gamma_{17}$\\\hline
$\Gamma_{19},\Gamma_{20}$,$\Gamma_{21}$&$\Gamma_{19}$\\\hline
\end{tabular}
\end{center}
\caption{Degenerated structures in the single field limit}
\end{table}
\newline One can now regard the single field case as an additional check. All terms involving $\varphi$ in the denominator should either vanish due to an accompanying factor of $(N-1)$ or should cancel exactly with other contributions coming from the degenerated structures. Indeed, in the single field limit there is no explicit dependence on $\varphi$ at all.\\
Still one may be worried about the terms with powers of $\varphi$ in the denominator in the multiplet case, since there is no reason why $\varphi=0$ should be a special value and $U(0)$, $V(0)$, $G(0)$ should be perfectly regular at $\varphi=0$. However, one can show that these inverse powers of $\varphi$ are just an artefact of our notation. Instead of considering the $O(N)$-invariant quantity $\varphi^2$, we focus on the field $\varphi=\sqrt{\Phi_{a}\Phi^{a}}$, since it corresponds to the inflaton in the cosmological application of the Higgs inflation and it faciliates the comparison with single field results. Due to the $O(N)$ symmetry all ``well-behaved'' potentials $U,V,G$ should \emph{effectively} depend on $\varphi^2$. Defining 
\begin{align}
\dot{Z}\equiv\frac{\partial Z(\varphi^2)}{\partial \varphi^2}\;, 
\end{align}
one can express the coefficients $\alpha_{i}$ in terms of the ``dot'' derivatives. The expression $s$ remains unchanged, but all derivatives $Z',Z'',Z'''$ are replaced by $\dot{Z}, \ddot{Z},\dddot{Z}$. The conversion formulae between the derivatives are given by
\begin{align}
Z'&=2\,\varphi\, \dot{Z}\;,\\
Z''&=4\,\varphi^{2}\, \ddot{Z}+2\,\dot{Z}\;, \\
Z'''&=8\,\varphi^{3}\, \dddot{Z}+12\, \varphi\, \ddot{Z}\;.
\end{align}
In that way, $\varphi$'s are generated in the numerator which then can cancel the $\varphi$ in the denominator.
Moreover, the inverse powers of $\varphi$ arising from the $n_{a}=\frac{\Phi_{a}}{\varphi}$ contained in some strutures $\Gamma_{i}$ must also at least be compensated by powers of $\varphi$ in the numerator of $\alpha_{i}$. Ultimatively, there can only remain even powers of $\varphi$ in the numerator.

\subsection{Comparsion with Known Results}

\subsubsection{Single Field \& Einstein Frame}
The comparison with \cite{we-renorm} corresponds to the Einstein frame result of the quantum corrections in the single field case. This means that in addition to the single field limit described above we set $U=G=1$. Our result obtained in this limit coincides with that obtained in \cite{we-renorm} up to a different coefficient in $V^2$ -- a misprint in \cite{we-renorm} which was already discovered and mentioned in \cite{we2}.\\
From the coincidence with this result, the coincidence with the well-known result of \cite{HV} for the case $V=0$ follows automatically.
 
\subsubsection{Multiplet in a Cosmological Setup}
In the context of the RG improved Higgs inflation scenario \cite{BKKSS} the beta-functions for the non-minimal coupling $\xi$ and the Higgs self-coupling $\lambda$ were derived by explicit calculations of $U_{1\text{-loop}}$ and $V_{1\text{-loop}}$. The tree level values of the couplings in this model are given by 
\begin{flalign}
U_{\text{tree}}(\varphi)=&\frac{1}{2}(M_{\text{P}}^2+\xi\varphi^2)\label{treelevelchargeU}\;,\\
G_{\text{tree}}(\varphi)=&1\label{treelevelchargeG}\;,\\
V_{\text{tree}}(\varphi)=&\frac{1}{4}\lambda(\varphi^2-\nu^2)^2\label{treelevelchargeV}\;.
\end{flalign}
Here $M_{\text{P}}$ is the reduced Planck mass, and $\nu$ is a symmetry breaking scale.
In the calculations of \cite{BKKSS} a multiplet $\Phi^{a}$ with a constant background $\nabla_{\mu}\bar{\Phi}^{a}=0$ (leading to vanishing $\hat{\cal{R}}_{\mu\nu}$) was assumed. Furthermore, the contribution of graviton loops was neglected (corresponding to considering only the scalar-scalar sector of $\hat{P}$). Finally, an expansion in $\xi^{-1}$ was made, which is justified by the assumption of a strong non-minimal coupling $\xi\simeq10^{4}-10^{5}$. 
Applying all these approximations to our general result reproduces exactly the result derived in the Appendix of \cite{BKKSS}.

\subsubsection{Single Field \& Jordan Frame}
Another important source for a cross-check is the result of Shapiro and Takata \cite{shapiro}. They did similar calculations for a single scalar field.
To confront our general result in the limiting case of a single field with the result of \cite{shapiro} we have to bring our result into a form that is suitable for comparison. We list the following relations in order to convert between our formalism and the one used in \cite{shapiro}
\begin{align}
&\phi\leftrightarrow\varphi\;,\\
 &A\leftrightarrow-\frac{1}{2}\,G,\;B\leftrightarrow U,\;C\leftrightarrow-V\;,\\
& A_{1}\leftrightarrow-\frac{1}{2}\,G',A_{2}\leftrightarrow-\frac{1}{2}\,G'', \text{ etc.}\\
&X\equiv2AB-3B_{1}^2\leftrightarrow-(GU+3U'^2)=-\frac{U}{s}\;.
\end{align}
In addition, the authors of \cite{shapiro} have chosen to express the $R^2$ and $R^{\mu\nu}R_{\mu\nu}$ contributions in terms of the Weyl tensor. Using the definition of the Weyl tensor 
\begin{align}
C_{\alpha\beta\gamma\delta}=R_{\alpha\beta\gamma\delta}-(g_{\alpha[\gamma}\,R_{\delta]\beta}-g_{\beta[\gamma}\,R_{\delta]\alpha})+\frac{1}{3}\,g_{\alpha[\gamma}g_{\delta]\beta}\,R
\end{align}
and the Gau\ss-Bonnet identity in four dimensions, we obtain for a general function $F(\phi)$
\begin{align}
F\, C^{\alpha\beta\gamma\delta}C_{\alpha\beta\gamma\delta}=&\,F\,\Big(\frac{1}{3}R^2-2R^{\alpha\beta}R_{\alpha\beta}+R^{\alpha\beta\gamma\delta}R_{\alpha\beta\gamma\delta}\Big)\nonumber\\
=&-\frac{2}{3}\,F\,R^2+2\,F\,R^{\alpha\beta}R_{\alpha\beta}\;.
\label{Weylsquared}
\end{align}
In the single field limit the number of different structures which do arise in the effective action reduces to eleven (see TABLE I).
Comparing our one-loop result in the single field limit with the one calculated in \cite{shapiro}, we find coincidence for the coefficients 
\begin{align}
c_{12}&\leftrightarrow \alpha_{1}\;,\\
c_{7}&\leftrightarrow \alpha_{2}\;,\\
c_{11}&\leftrightarrow \frac{1}{2}\alpha_{3}\;,\\
2c_{w}&\leftrightarrow \alpha_{5}\;,\\
 c_{r}-\frac{2}{3}c_{w}&\leftrightarrow \alpha_{6}\;,\\
c_{10}&\leftrightarrow \alpha_{18}\;.
\end{align}
Differences remain for all other coefficients. To trace back the origin of these discrepancies, we first have reduced our input of the MathTensor code (the second variation) to the single field case and compared it with the input used in \cite{shapiro}. The authors of \cite{shapiro} have used the convention
\begin{align}
\omega^{A}\equiv \delta\Psi^{A}=\left(\begin{array}{c}
                \bar{h}_{\mu\nu}\\
		h\\
		\sigma 
                \end{array}\right)\;,
\end{align}
with $\delta g_{\mu\nu}\equiv h_{\mu\nu},\; \delta\phi\equiv\sigma$, $\bar{h}_{\mu\nu}=h_{\mu\nu}-\frac{1}{4}g^{\mu\nu}h_{\mu\nu}$ and $h=g^{\mu\nu}h_{\mu\nu}$. We did not split $h_{\mu\nu}$ in the traceless part $\bar{h}_{\mu\nu}$ and its trace $h$. To compare their operator $F_{AB}$ ($3\times3$ matrix) with ours ($2\times2$ matrix) it is useful to expand the quadratic forms of the composing parts $\omega^{A}\tilde{C}_{AB}\Box\,\omega^{B}$, $\omega^{A}\Gamma_{AB}^{\mu}\nabla_{\mu}\omega^{B}$, $\omega^{A}W_{AB}\,\omega^{B}$ and compare the resulting scalars.
Both expressions coincide up to a factor of 
\begin{align}
 \frac{1}{4}\,h\,A_{1}\,(\nabla\phi)^2\,\sigma
 \label{input difference}
\end{align}
present in \cite{shapiro}. Direct recalculations starting with the action of \cite{shapiro} confirmed our result that this factor should indeed be absent. The origin of this is a cancellation due to contributions coming from partial integration.
This can be seen easily by calculating the mixed variations and concentrating on the trace part of the metric perturbations.\\
Since our input and the one of \cite{shapiro} coincide up to the term (\ref{input difference}), we can check if the differences in the final one-loop results all vanish when setting $A_{1}$ to zero. This is not the case, although the differences vanish e.g.~for the $c_{6}\leftrightarrow \alpha_{7}$ coefficient in this case. We have programmed a new MathTensor algorithm to calculate the single scalar field case directly with MathTensor. We obtained the same result as in the single field limit of our general $O(N)$ result. To further investigate the remaining differences (independent of $A_{1}$) we  considered the coefficient with the biggest deviation $c_8$. To faciliate the calculations by hand, we limited ourselves to the special case $U=B=1$ and focused on the $G''=-\frac{1}{2}A_{2}$ contributions in $c_{8}$. In our calculations (also repeated with MathTensor) these contributions are absent, but in \cite{shapiro} there remains a contribution $G''$. 

To summarize, our results are in agreement with those from paper \cite{shapiro} for the most important structures such as 
$V_{1\text{-loop}}, U_{1\text{-loop}}, G_{1\text{-loop}}$. As far as concerned other structures, 
we have found one source of discrepancies: it is connected with the presence of the  factor (\ref{input difference}) in the input 
of \cite{shapiro} while this term is absent in our input. There still remain  some discrepancies, which cannot be reduced 
to this difference in the input. Here, our belief in the correctness of our results is based on some additional cross-checks, including the calculations made by hands for some limiting cases. 

\section{Conclusion and outlook}
By applying the generalized Schwinger-DeWitt technique \cite{DeWitt,Bar-Vil} and using the MathTensor package for Mathematica \cite{Mathtensor}, we have calculated the divergent part of the one-loop effective action for a multiplet of scalar fields non-minimally coupled to gravity in a closed form. All the calculations were done in the Jordan frame. In the next paper 
\cite{we-future} we will compare our Jordan frame results with those by first performing a transformation to the Einstein frame at the classical level, then calculating the effective action in the Einstein frame, and finally performing the inverse transformation back to the Jordan frame. To that end we will present the transformation rules for the transition between Jordan and Einstein frame for the more general case of an $O(N)$ multiplet of scalar fields. We will see that the results obtained by calculating quantum corrections in the two different frames are different. This suggests that for the correct calculation of the quantum corrections to physical quantities one should always 
use the same frame. Of course, this does not yet answer the question which of the both frames is the physical one and should be used. We will investigate this problem in \cite{we-future}. The third paper of the series \cite{we-future1} will be devoted to several cosmological applications of the results 
obtained in the present paper. Basically, it will concern the effect of suppression of the contributions of the Higgs 
propagators to the quantum corrections, which is caused by the non-minimal coupling between scalar fields and gravity, and the question of the limits of the applicability of the perturbative expansion in cosmology.
\section*{Acknowledgments}
We are grateful to A. O. Barvinsky and C. Kiefer for numerous fruitful discussions. We also want to thank I. L. Shapiro for a friendly and helpful correspondence and an anonymous referee for useful advice. 
A. K. acknowledges support by the grant 436 RUS 17/3/07 of the
German Science Foundation (DFG) and by the University of Cologne for his visits to the Cologne Institute for Theoretical Physics. He was
also partially supported by the RFBR grant 11-02-00643. 
 The work of C. F. S.
was supported by the Villigst Foundation.
\newpage
\onecolumngrid
\appendix\section{Transfer Equations}
Not all structures appearing in the calculations are independent. Neglecting surface terms and making use of the Bianchi identities, one can convert certain structures into others via integration by parts. In such a way one can reduce the number of different structures in (\ref{effectiveaction}) to a minimum. The ``transfer equations'' below describe explicitly how the contributions of the dependent structures are distributed among the minimal set of independent structures:
\begin{align} 
&\quad F\,\Phi^{a\;\;\mu}_{;\,\mu}\,n_{a}\rightarrow -F'\,\Gamma_{4}-\frac{F}{\varphi}\,(\Gamma_{3}-\Gamma_{4})\label{transferequation1}\\\nonumber\\
&\quad F\,\Phi^{a}_{;\,\mu\nu}\Phi_{a}^{;\,\mu\nu}\rightarrow F'\,\Gamma_{21}-F(\Gamma_{7}-\Gamma_{17})+\frac{F'}{\varphi}\,\Gamma_{15}-\frac{1}{2}\Big(\frac{F'}{\varphi}-F''\Big)\,\Gamma_{14}+\frac{1}{2}F'\,\Gamma_{19}-\frac{1}{2}\frac{F'}{\varphi}\,\Gamma_{12}\\\nonumber\\
&\quad F\,\Phi^{a}_{;\,\mu\nu}\,n_{a}\Phi^{b;\,\mu\nu}\,n_{b}\rightarrow\Big(4\frac{F}{\varphi^2}-\frac{5}{2}\,\frac{F'}{\varphi}+\frac{1}{2}\,F''\Big)\,\Gamma_{13}+\Big(2\,\frac{F'}{\varphi}-4\,\frac{F}{\varphi^2}\Big)\,\Gamma_{16}+\Big(\frac{3}{2}\,F'-3\,\frac{F}{\varphi}\Big)\,\Gamma_{20}\nonumber\\
&\quad\quad\quad\quad\quad\quad\quad\quad\quad\quad+\Big(\frac{1}{2}\,\frac{F'}{\varphi}-\frac{F}{\varphi^2}\Big)\,\Gamma_{14}+2\,\frac{F}{\varphi}\,\Gamma_{21}-F\,\Gamma_{8}+\frac{F}{\varphi}\,\Gamma_{19}+F\,\Gamma_{18}+\frac{F}{\varphi^2}\,\Gamma_{15}\\\nonumber\\
&\quad F\,\Phi^{a}_{;\,\mu\nu}\,n_{a}\,\Phi^{b,\,\mu}\,\Phi_{b}^{,\,\nu}\rightarrow-\frac{1}{2}\,\Big(\frac{F}{\varphi}-F'\Big)\,\Gamma_{14}+\Big(\frac{F}{\varphi}-F'\Big)\,\Gamma_{16}-F\,\Gamma_{21}+\frac{1}{2}\,F\,\Gamma_{19}+\frac{1}{2}\,\frac{F}{\varphi}\,\Gamma_{12}-\frac{F}{\varphi}\,\Gamma_{15}\\\nonumber\\
&\quad F\,\Phi^{a}_{;\,\mu\nu}\Phi^{,\,\mu}_{a}\,\Phi^{,\,\nu}_{b}\,n^{b}\rightarrow \frac{1}{2}\,\Big(\frac{F}{\varphi}-F'\Big)\,\Gamma_{14}-\frac{F}{2}\,\Gamma_{19}-\frac{1}{2}\,\frac{F}{\varphi}\,\Gamma_{12}\\\nonumber\\
&\quad F\,\Phi^{a}_{;\,\mu\nu}\,n_{a}\,\Phi_{b}^{,\,\mu}\,n^{b}\,\Phi_{c}^{,\,\nu}\,n^{c}\rightarrow\frac{1}{2}\,\Big(3\,\frac{F}{\varphi}-F'\Big)\,\Gamma_{13}-\frac{F}{\varphi}\,\Gamma_{16}-\frac{1}{2}\,F\,\Gamma_{20}-\frac{1}{2}\,\frac{F}{\varphi}\,\Gamma_{14}\\\nonumber\\
&\quad F\,R^{\mu\nu}\,\Phi^{a}_{;\,\mu\nu}\,n_{a}\,\rightarrow \Big(\frac{F}{\varphi}-F'\Big)\,\Gamma_{8}-\frac{1}{2}\,\Big(\frac{F}{\varphi}-F'\Big)\,\Gamma_{10}-\frac{F}{\varphi}\,\Gamma_{7}+\frac{1}{2}\,F\,\Gamma_{11}+\frac{1}{2}\,\frac{F}{\varphi}\,\Gamma_{9}\\\nonumber\\
 &\quad F\,R^{\mu\nu\rho\sigma}R_{\mu\nu\rho\sigma}\rightarrow4\,F\,\Gamma_{5}-F\,\Gamma_{6}\;.
 \label{transfer equations}
\end{align}
The last equation is a topological invariant, the Gau\ss\,-Bonnet identity. In the single field limit (see TABLE I) the other seven transfer equations reduce to the three reduction formulae given in \cite{shapiro}. In principle, even more scalar invariants composed of different scalar contractions between derivatives $\nabla_{\mu}$ and field variables $(g_{\mu\nu},\,\Phi^{a})$ (containing up to four derivatives) can appear at the one-loop level, but they do not occur in our calculations using the Schwinger-DeWitt algorithm.\\
\section{Gradient Structures}
 As already explained in the main text, the coefficients $\alpha_{i},\,i=12,...,21$, all corresponding to the symbolical structure $\partial^4\Phi\Phi\Phi\Phi$, are additionally suppressed compared to (\ref{V1loop})-(\ref{alphaeleven}) and thus less important. However, for the sake of completeness we also want to list the remaining coefficients in a closed form.\\\\

The coefficient $ \alpha_{12}$ in front of the structure $(\Phi^{a}_{,\,\mu}\Phi_{a}^{,\,\mu})^2$ :
\begin{align}
 \alpha_{12}=&\; s^2
   \Bigg(\frac{25 \left(U'\right)^8}{48 G^2 \varphi ^4
   U^4}+\frac{3 \left(U'\right)^8}{16
   U^6}-\frac{\left(U'\right)^7}{4 G \varphi ^3 U^4}-\frac{3
   \left(U'\right)^7}{8 \varphi  U^5}+\frac{5 G'
   \left(U'\right)^6}{4 G^2 \varphi ^3 U^3}-\frac{11
   \left(U'\right)^6}{72 G \varphi ^4 U^3}-\frac{9
   \left(U'\right)^6}{8 \varphi ^2 U^4}-\frac{13 G'
   \left(U'\right)^5}{12 G \varphi ^2 U^3}-\frac{G'
   \left(U'\right)^5}{2 U^4}\nonumber\\
&\;-\frac{3 U'' \left(U'\right)^5}{4 G
   \varphi ^3 U^3}+\frac{35 \left(U'\right)^5}{12 \varphi ^3
   U^3}+\frac{13 \left(G'\right)^2 \left(U'\right)^4}{24 G^2
   \varphi ^2 U^2}-\frac{G' \left(U'\right)^4}{G \varphi ^3
   U^2}+\frac{15 G' \left(U'\right)^4}{8 \varphi  U^3}+\frac{3
   G'' \left(U'\right)^4}{8 U^3}-\frac{695
   \left(U'\right)^4}{432 \varphi ^4
   U^2}+\frac{\left(G'\right)^2 \left(U'\right)^3}{8 G \varphi 
   U^2}\nonumber\\
&\;-\frac{47 G' \left(U'\right)^3}{18 \varphi ^2
   U^2}-\frac{3 G'' \left(U'\right)^3}{4 \varphi  U^2}+\frac{3
   G' U'' \left(U'\right)^3}{4 G \varphi ^2 U^2}-\frac{U''
   \left(U'\right)^3}{4 \varphi ^3 U^2}-\frac{\left(G'\right)^3
   \left(U'\right)^2}{4 G^2 \varphi  U}+\frac{139
   \left(G'\right)^2 \left(U'\right)^2}{72 G \varphi ^2
   U}+\frac{\left(G'\right)^2 \left(U'\right)^2}{8 U^2}+\frac{19
   G' \left(U'\right)^2}{36 \varphi ^3 U}\nonumber\\
&\;+\frac{3 G' G''
   \left(U'\right)^2}{4 G \varphi  U}+\frac{\left(G'\right)^3
   U'}{6 G U}-\frac{\left(G'\right)^2 U'}{12 \varphi 
   U}-\frac{G' G'' U'}{4 U}+\frac{G' U'' U'}{4 \varphi ^2
   U}+\frac{\left(G'\right)^4}{48 G^2}-\frac{7
   \left(G'\right)^3}{24 G \varphi }+\frac{11
   \left(G'\right)^2}{24 \varphi
   ^2}+\frac{\left(G''\right)^2}{8}-\frac{\left(G'\right)^2
   G''}{8 G}\nonumber\\
&\;+\frac{G' G''}{2 \varphi }\Bigg)+s \Bigg(-\frac{25
   \left(U'\right)^6}{72 G^2 \varphi ^4
   U^3}-\frac{\left(U'\right)^6}{8
   U^5}+\frac{\left(U'\right)^5}{6 G \varphi ^3 U^3}-\frac{37
   \left(U'\right)^5}{8 \varphi  U^4}+\frac{5 G'
   \left(U'\right)^4}{12 G^2 \varphi ^3 U^2}-\frac{43
   \left(U'\right)^4}{216 G \varphi ^4 U^2}+\frac{21
   \left(U'\right)^4}{4 \varphi ^2 U^3}+\frac{13 G'
   \left(U'\right)^3}{36 G \varphi ^2 U^2}\nonumber\\
&\;+\frac{11 G'
   \left(U'\right)^3}{12 U^3}+\frac{U'' \left(U'\right)^3}{2 G
   \varphi ^3 U^2}+\frac{3 U'' \left(U'\right)^3}{2 \varphi 
   U^3}-\frac{107 \left(U'\right)^3}{36 \varphi ^3 U^2}+\frac{77
   \left(G'\right)^2 \left(U'\right)^2}{72 G^2 \varphi ^2
   U}+\frac{15 G' \left(U'\right)^2}{4 G \varphi ^3 U}-\frac{15
   G' \left(U'\right)^2}{8 \varphi  U^2}-\frac{G''
   \left(U'\right)^2}{8 U^2}\nonumber\\
&\;-\frac{3 U''
   \left(U'\right)^2}{\varphi ^2
   U^2}-\frac{\left(U'\right)^2}{12 \varphi ^4 U}+\frac{5
   \left(G'\right)^2 U'}{6 G \varphi  U}+\frac{13 G' U'}{6
   \varphi ^2 U}-\frac{G' U'' U'}{4 G \varphi ^2 U}+\frac{U''
   U'}{12 \varphi ^3 U}-\frac{\left(G'\right)^3}{6 G^2 \varphi
   }-\frac{13 \left(G'\right)^2}{12 G \varphi ^2}-\frac{G'
   G''}{4 G \varphi }+\frac{G' U''}{2 \varphi  U}\Bigg)\nonumber\\
&+\frac{25 \left(U'\right)^4}{432 G^2 \varphi ^4
   U^2}+\frac{\left(U'\right)^4}{48 U^4}+\frac{59
   \left(U'\right)^3}{36 G \varphi ^3 U^2}+\frac{19
   \left(U'\right)^3}{12 \varphi  U^3}-\frac{G'
   \left(U'\right)^2}{9 G^2 \varphi ^3 U}+\frac{5
   \left(U'\right)^2}{12 G \varphi ^4 U}+\frac{17
   \left(U'\right)^2}{3 \varphi ^2 U^2}-\frac{4 G' U'}{3 G
   \varphi ^2 U}-\frac{G' U'}{4 U^2}-\frac{5 U'' U'}{12 G
   \varphi ^3 U}\nonumber\\
&\;-\frac{U'' U'}{2 \varphi  U^2}-\frac{5 U'}{3
   \varphi ^3 U}+\frac{13 G U'}{6 \varphi  U^2}+\frac{(N-1)
   \left(G'\right)^2}{2 G^2 \varphi ^2}-\frac{5
   \left(G'\right)^2}{12 G^2 \varphi ^2}-\frac{7 G'}{6 \varphi 
   U}+\frac{5 U''}{3 \varphi ^2 U}-\frac{G^2}{4 U^2}
\end{align}
\\

The coefficient $ \alpha_{13}$ in front of the structure $(\Phi^{a}_{,\,\mu}n_{a}\Phi^{b\,,\,\mu}n_{b})^2$ :
\begin{align}
 \alpha_{13}=&\; s^4\Bigg(\frac{81 \left(U'\right)^{12}}{32 U^8}-\frac{81 U''
   \left(U'\right)^{10}}{4 U^7}-\frac{27 G' \left(U'\right)^9}{8
   U^6}+\frac{243 \left(U''\right)^2 \left(U'\right)^8}{4
   U^6}+\frac{81 G' U'' \left(U'\right)^7}{4 U^5}-\frac{81
   \left(U''\right)^3 \left(U'\right)^6}{U^5}+\frac{27
   \left(G'\right)^2 \left(U'\right)^6}{16 U^4}\nonumber\\
&\;-\frac{81 G'
   \left(U''\right)^2 \left(U'\right)^5}{2 U^4}+\frac{81
   \left(U''\right)^4 \left(U'\right)^4}{2 U^4}-\frac{27
   \left(G'\right)^2 U'' \left(U'\right)^4}{4 U^3}-\frac{3
   \left(G'\right)^3 \left(U'\right)^3}{8 U^2}+\frac{27 G'
   \left(U''\right)^3 \left(U'\right)^3}{U^3}\nonumber\\
&\;+\frac{27
   \left(G'\right)^2 \left(U''\right)^2 \left(U'\right)^2}{4
   U^2}+\frac{3 \left(G'\right)^3 U'' U'}{4
   U}+\frac{\left(G'\right)^4}{32}\Bigg)+ s^3\Bigg(-\frac{3
   \left(U'\right)^{10}}{2 G \varphi ^2 U^6}+\frac{45
   \left(U'\right)^{10}}{8 U^7}-\frac{81 \left(U'\right)^9}{8
   \varphi  U^6}-\frac{9 G' \left(U'\right)^8}{4 G \varphi 
   U^5}\nonumber\\
&\;+\frac{19 U'' \left(U'\right)^8}{4 G \varphi ^2
   U^5}+\frac{27 U'' \left(U'\right)^8}{8 U^6}+\frac{25
   \left(U'\right)^8}{4 \varphi ^2 U^5}+\frac{G'
   \left(U'\right)^7}{G \varphi ^2 U^4}-\frac{15 G'
   \left(U'\right)^7}{4 U^5}+\frac{81 U'' \left(U'\right)^7}{2
   \varphi  U^5}+\frac{5 \left(G'\right)^2 \left(U'\right)^6}{8
   G U^4}+\frac{155 \left(U''\right)^2 \left(U'\right)^6}{8 G
   \varphi ^2 U^4}\nonumber\\
&\;-\frac{351 \left(U''\right)^2
   \left(U'\right)^6}{4 U^5}+\frac{39 G' \left(U'\right)^6}{8
   \varphi  U^4}+\frac{9 G'' \left(U'\right)^6}{8 U^4}+\frac{3
   G' U'' \left(U'\right)^6}{G \varphi  U^4}-\frac{305 U''
   \left(U'\right)^6}{12 \varphi ^2 U^4}+\frac{3
   \left(G'\right)^2 \left(U'\right)^5}{2 G \varphi 
   U^3}-\frac{81 \left(U''\right)^2 \left(U'\right)^5}{2 \varphi
    U^4}\nonumber\\
&\;-\frac{25 G' \left(U'\right)^5}{6 \varphi ^2
   U^3}+\frac{15 G' U'' \left(U'\right)^5}{4 G \varphi ^2
   U^3}-\frac{39 G' U'' \left(U'\right)^5}{4 U^4}+\frac{261
   \left(U''\right)^3 \left(U'\right)^4}{2 U^4}+\frac{29
   \left(G'\right)^2 \left(U'\right)^4}{24 U^3}+\frac{93 G'
   \left(U''\right)^2 \left(U'\right)^4}{4 G \varphi 
   U^3}\nonumber\\
&\;+\frac{803 \left(U''\right)^2 \left(U'\right)^4}{24
   \varphi ^2 U^3}+\frac{\left(G'\right)^2 U''
   \left(U'\right)^4}{G U^3}-\frac{8 G' U''
   \left(U'\right)^4}{\varphi  U^3}-\frac{9 G'' U''
   \left(U'\right)^4}{2 U^3}+\frac{\left(G'\right)^2
   \left(U'\right)^3}{8 \varphi  U^2}+\frac{39 G'
   \left(U''\right)^2 \left(U'\right)^3}{U^3}\nonumber\\
&\;-\frac{3 G' G''
   \left(U'\right)^3}{4 U^2}+\frac{9 \left(G'\right)^2 U''
   \left(U'\right)^3}{G \varphi  U^2}+\frac{41 G' U''
   \left(U'\right)^3}{4 \varphi ^2 U^2}-\frac{27
   \left(U''\right)^4 \left(U'\right)^2}{U^3}+\frac{3
   \left(G'\right)^3 \left(U'\right)^2}{4 G \varphi  U}+\frac{3
   \left(G'\right)^2 \left(U'\right)^2}{4 \varphi ^2 U}\nonumber\\
&\;-\frac{31
   \left(G'\right)^2 \left(U''\right)^2 \left(U'\right)^2}{8 G
   U^2}+\frac{13 G' \left(U''\right)^2 \left(U'\right)^2}{4
   \varphi  U^2}+\frac{9 G'' \left(U''\right)^2
   \left(U'\right)^2}{2 U^2}+\frac{41 \left(G'\right)^2 U''
   \left(U'\right)^2}{24 U^2}-\frac{\left(G'\right)^3 U'}{4
   U}-\frac{9 G' \left(U''\right)^3 U'}{U^2}\nonumber\\
&\;-\frac{3
   \left(G'\right)^3 U'' U'}{2 G U}+\frac{3 \left(G'\right)^2
   U'' U'}{2 \varphi  U}+\frac{3 G' G'' U'' U'}{2
   U}-\frac{\left(G'\right)^4}{8 G}+\frac{\left(G'\right)^3}{8
   \varphi }-\frac{13 \left(G'\right)^2 \left(U''\right)^2}{24
   U}+\frac{1}{8} \left(G'\right)^2 G''\Bigg)\nonumber\\
& +  s^2\Bigg(\frac{25 \left(U'\right)^8}{16 G^2 \varphi ^4
   U^4}-\frac{19 \left(U'\right)^8}{6 G \varphi ^2
   U^5}-\frac{189 \left(U'\right)^8}{16 U^6}-\frac{5
   \left(U'\right)^7}{4 G \varphi ^3 U^4}-\frac{123
   \left(U'\right)^7}{8 \varphi  U^5}+\frac{15 G'
   \left(U'\right)^6}{4 G^2 \varphi ^3 U^3}+\frac{G'
   \left(U'\right)^6}{4 G \varphi  U^4}-\frac{U''
   \left(U'\right)^6}{6 G \varphi ^2 U^4}\nonumber\\
&\;+\frac{273 U''
   \left(U'\right)^6}{8 U^5}+\frac{133 \left(U'\right)^6}{24 G
   \varphi ^4 U^3}-\frac{223 \left(U'\right)^6}{72 \varphi ^2
   U^4}-\frac{25 G' \left(U'\right)^5}{24 G \varphi ^2
   U^3}+\frac{85 G' \left(U'\right)^5}{8 U^4}-\frac{5 G''
   \left(U'\right)^5}{4 G \varphi  U^3}-\frac{15 G' U''
   \left(U'\right)^5}{4 G^2 \varphi ^2 U^3}-\frac{33 U''
   \left(U'\right)^5}{4 G \varphi ^3 U^3}\nonumber\\
&\;+\frac{87 U''
   \left(U'\right)^5}{4 \varphi  U^4}+\frac{3 U'''
   \left(U'\right)^5}{4 G \varphi ^2 U^3}-\frac{3 U'''   \left(U'\right)^5}{2 U^4}-\frac{113 \left(U'\right)^5}{12
   \varphi ^3 U^3}+\frac{\left(G'\right)^2
   \left(U'\right)^4}{G^2 \varphi ^2 U^2}-\frac{4
   \left(G'\right)^2 \left(U'\right)^4}{3 G U^3}-\frac{55
   \left(U''\right)^2 \left(U'\right)^4}{6 G \varphi ^2
   U^3}\nonumber\\
&\;-\frac{81 \left(U''\right)^2 \left(U'\right)^4}{8
   U^4}+\frac{5 G' \left(U'\right)^4}{2 G \varphi ^3
   U^2}+\frac{167 G' \left(U'\right)^4}{24 \varphi  U^3}-\frac{7
   G'' \left(U'\right)^4}{8 U^3}+\frac{45 G' U''
   \left(U'\right)^4}{4 G \varphi  U^3}+\frac{881 U''
   \left(U'\right)^4}{36 \varphi ^2 U^3}+\frac{889
   \left(U'\right)^4}{144 \varphi ^4 U^2}\nonumber\\
&\;-\frac{7
   \left(G'\right)^2 \left(U'\right)^3}{8 G \varphi 
   U^2}+\frac{9 \left(U''\right)^2 \left(U'\right)^3}{2 \varphi 
   U^3}+\frac{325 G' \left(U'\right)^3}{72 \varphi ^2
   U^2}-\frac{8 G'' \left(U'\right)^3}{3 \varphi  U^2}-\frac{9
   \left(G'\right)^2 U'' \left(U'\right)^3}{2 G^2 \varphi 
   U^2}-\frac{31 G' U'' \left(U'\right)^3}{2 G \varphi ^2
   U^2}-\frac{25 G' U'' \left(U'\right)^3}{U^3}\nonumber\\
&\;-\frac{13 G'' U''
   \left(U'\right)^3}{4 G \varphi  U^2}-\frac{11 U''
   \left(U'\right)^3}{4 \varphi ^3 U^2}-\frac{3 G' U'''   \left(U'\right)^3}{4 G \varphi  U^2}-\frac{6 U'' U'''   \left(U'\right)^3}{U^3}+\frac{U'''\left(U'\right)^3}{4
   \varphi ^2 U^2}-\frac{3 \left(G'\right)^3
   \left(U'\right)^2}{2 G^2 \varphi  U}-\frac{39
   \left(U''\right)^3 \left(U'\right)^2}{2 U^3}\nonumber\\
&\;-\frac{8
   \left(G'\right)^2 \left(U'\right)^2}{3 G \varphi ^2
   U}-\frac{19 \left(G'\right)^2 \left(U'\right)^2}{8
   U^2}-\frac{43 G' \left(U''\right)^2 \left(U'\right)^2}{4 G
   \varphi  U^2}-\frac{713 \left(U''\right)^2
   \left(U'\right)^2}{72 \varphi ^2 U^2}-\frac{31 G'
   \left(U'\right)^2}{12 \varphi ^3 U}-\frac{125
   \left(G'\right)^2 U'' \left(U'\right)^2}{24 G U^2}\nonumber\\
&\;-\frac{2 G'
   U'' \left(U'\right)^2}{3 \varphi  U^2}+\frac{7 G'' U''
   \left(U'\right)^2}{4 U^2}-\frac{G' U'''   \left(U'\right)^2}{2 U^2}-\frac{\left(G'\right)^3 U'}{8 G
   U}-\frac{23 \left(G'\right)^2 U'}{12 \varphi  U}+\frac{17 G'
   \left(U''\right)^2 U'}{2 U^2}-\frac{G' G'' U'}{4 U}\nonumber\\
&\;+\frac{3
   \left(G'\right)^3 U'' U'}{4 G^2 U}-\frac{4 G' U'' U'}{3
   \varphi ^2 U}+\frac{3 G' G'' U'' U'}{2 G U}+\frac{23 G'' U''
   U'}{12 \varphi  U}-\frac{G' U'''U'}{4 \varphi  U}+\frac{3
   \left(G'\right)^4}{16 G^2}+\frac{9 \left(U''\right)^4}{2
   U^2}+\frac{\left(G'\right)^3}{8 G \varphi }+\frac{5
   \left(G'\right)^2}{8 \varphi
   ^2}\nonumber\\
&\;+\frac{\left(G''\right)^2}{8}+\frac{13 \left(G'\right)^2
   \left(U''\right)^2}{24 G U}-\frac{5 G' \left(U''\right)^2}{2
   \varphi  U}-\frac{3 G'' \left(U''\right)^2}{2
   U}+\frac{\left(G'\right)^2 G''}{8 G}+\frac{3 G' G''}{4
   \varphi }+\frac{\left(G'\right)^2 U''}{2 U}\Bigg)\nonumber\\
&+s\Bigg(-\frac{25 \left(U'\right)^6}{24 G^2 \varphi ^4
   U^3}-\frac{35 \left(U'\right)^6}{9 G \varphi ^2 U^4}+\frac{61
   \left(U'\right)^6}{8 U^5}+\frac{7 G' \left(U'\right)^5}{8 G^2
   \varphi ^2 U^3}-\frac{37 \left(U'\right)^5}{6 G \varphi ^3
   U^3}+\frac{189 \left(U'\right)^5}{8 \varphi  U^4}+\frac{5
   \left(G'\right)^2 \left(U'\right)^4}{8 G^3 \varphi ^2
   U^2}-\frac{3 G' \left(U'\right)^4}{8 G^2 \varphi ^3
   U^2}\nonumber\\
&\;-\frac{11 G' \left(U'\right)^4}{12 G \varphi 
   U^3}-\frac{5 G'' \left(U'\right)^4}{8 G^2 \varphi ^2
   U^2}+\frac{16 U'' \left(U'\right)^4}{9 G \varphi ^2
   U^3}-\frac{227 U'' \left(U'\right)^4}{8 U^4}-\frac{79
   \left(U'\right)^4}{72 G \varphi ^4 U^2}+\frac{521
   \left(U'\right)^4}{27 \varphi ^2 U^3}-\frac{5
   \left(G'\right)^2 \left(U'\right)^3}{2 G^2 \varphi 
   U^2}-\frac{56 G' \left(U'\right)^3}{9 G \varphi ^2
   U^2}\nonumber\\
&\;-\frac{13 G' \left(U'\right)^3}{2 U^3}-\frac{19 G''
   \left(U'\right)^3}{12 G \varphi  U^2}+\frac{G' U''
   \left(U'\right)^3}{G^2 \varphi ^2 U^2}+\frac{3 U''
   \left(U'\right)^3}{2 G \varphi ^3 U^2}-\frac{95 U''
   \left(U'\right)^3}{2 \varphi  U^3}-\frac{U'''   \left(U'\right)^3}{2 G \varphi ^2 U^2}+\frac{3 U'''   \left(U'\right)^3}{U^3}+\frac{101 \left(U'\right)^3}{36
   \varphi ^3 U^2}\nonumber\\
&\;+\frac{3 \left(G'\right)^3
   \left(U'\right)^2}{4 G^3 \varphi  U}+\frac{17
   \left(G'\right)^2 \left(U'\right)^2}{8 G^2 \varphi ^2
   U}+\frac{\left(G'\right)^2 \left(U'\right)^2}{4 G
   U^2}+\frac{101 \left(U''\right)^2 \left(U'\right)^2}{72 G
   \varphi ^2 U^2}+\frac{53 \left(U''\right)^2
   \left(U'\right)^2}{2 U^3}+\frac{11 G' \left(U'\right)^2}{8 G
   \varphi ^3 U}-\frac{433 G' \left(U'\right)^2}{36 \varphi 
   U^2}\nonumber\\
&\;-\frac{41 G'' \left(U'\right)^2}{24 G \varphi ^2
   U}+\frac{G'' \left(U'\right)^2}{U^2}-\frac{17 G' U''
   \left(U'\right)^2}{6 G \varphi  U^2}-\frac{21 U''
   \left(U'\right)^2}{4 \varphi ^2 U^2}+\frac{3 U'''   \left(U'\right)^2}{\varphi  U^2}+\frac{\left(U'\right)^2}{4
   \varphi ^4 U}+\frac{3 \left(G'\right)^3 U'}{8 G^2 U}+\frac{17
   \left(G'\right)^2 U'}{12 G \varphi  U}\nonumber\\
&\;+\frac{3
   \left(U''\right)^2 U'}{2 \varphi  U^2}-\frac{17 G' U'}{12
   \varphi ^2 U}+\frac{G' G'' U'}{4 G U}-\frac{G'' U'}{2 \varphi
    U}+\frac{3 \left(G'\right)^2 U'' U'}{2 G^2 \varphi 
   U}+\frac{17 G' U'' U'}{4 G \varphi ^2 U}+\frac{27 G' U''
   U'}{2 U^2}+\frac{13 G'' U'' U'}{12 G \varphi  U}-\frac{5 U''
   U'}{12 \varphi ^3 U}\nonumber\\
&\;+\frac{G' U'''U'}{4 G \varphi 
   U}-\frac{U'' U'''U'}{U^2}-\frac{U'''U'}{12 \varphi ^2
   U}-\frac{\left(G'\right)^4}{8 G^3}-\frac{\left(G'\right)^3}{8
   G^2 \varphi }-\frac{\left(U''\right)^3}{2 U^2}+\frac{23
   \left(G'\right)^2}{24 U}-\frac{\left(G'\right)^2}{4 G \varphi
   ^2}-\frac{\left(G''\right)^2}{4 G}+\frac{G'
   \left(U''\right)^2}{G \varphi  U}+\frac{\left(U''\right)^2}{6
   \varphi ^2 U}\nonumber\\
&\;-\frac{\left(G'\right)^2 G''}{8 G^2}-\frac{G'
   G''}{2 G \varphi }+\frac{\left(G'\right)^2 U''}{2 G
   U}+\frac{25 G' U''}{12 \varphi  U}-\frac{G'
   U'''}{U}\Bigg)\nonumber\\
& +\frac{25 \left(U'\right)^4}{144 G^2 \varphi ^4
   U^2}+\frac{46 \left(U'\right)^4}{27 G \varphi ^2
   U^3}+\frac{85 \left(U'\right)^4}{8 U^4}-\frac{\left(G'\right)^3}{4
   G^3 \varphi }-\frac{7 G' \left(U'\right)^3}{24 G^2 \varphi ^2
   U^2}+\frac{133 \left(U'\right)^3}{36 G \varphi ^3
   U^2}+\frac{3 \left(U'\right)^3}{4 \varphi  U^3}-\frac{23
   \left(G'\right)^2}{24 G U}-\frac{\left(G'\right)^2}{2 G^2 \varphi
   ^2}\nonumber\\
&\;-\frac{5 \left(G'\right)^2 \left(U'\right)^2}{24 G^3
   \varphi ^2 U}-\frac{7 G' \left(U'\right)^2}{24 G^2 \varphi ^3
   U}-\frac{89 G' \left(U'\right)^2}{36 G \varphi 
   U^2}-\frac{\left(U'\right)^2}{4 G \varphi ^4 U}+\frac{3
   \left(U'\right)^2}{2 \varphi ^2 U^2}-\frac{\left(U''\right)^2}{6 G
   \varphi ^2 U}+\frac{9 \left(U''\right)^2}{2
   U^2}-\frac{\left(G'\right)^2 U'}{2 G^2 \varphi  U}-\frac{19
   G' U'}{12 G \varphi ^2 U}\nonumber\\
&\;+\frac{5 \left(U'\right)^2 G''}{24 G^2 \varphi ^2
   U}+\frac{U' G''}{2 G
   \varphi  U}-\frac{9 \left(U'\right)^2 U''}{4 G \varphi ^2
   U^2}-\frac{3 \left(U'\right)^2 U''}{4 U^3}+\frac{11 G'
   U''}{12 G \varphi  U}+\frac{G' U' U''}{12 G^2 \varphi ^2
   U}+\frac{5 U' U''}{12 G \varphi ^3 U}-\frac{6 U' U''}{\varphi
    U^2}+\frac{U' U'''}{12 G \varphi ^2 U}\nonumber\\
&\;-\frac{U'
   U'''}{U^2}+\frac{(N-1) \left(G'\right)^2}{8
   G^2 \varphi ^2}+\frac{(N-1)
   \left(G'\right)^3}{8 G^3 \varphi }+\frac{(N-1) \left(G'\right)^4}{32
   G^4}-\frac{(N-1) G' G''}{4 G^2 \varphi }-\frac{(N-1) \left(G'\right)^2
   G''}{8 G^3}+\frac{(N-1)
   \left(G''\right)^2}{8 G^2}
\end{align}
\\

The coefficient $ \alpha_{14}$ in front of the structure $(\Phi^{a}_{,\,\mu}\Phi_{a}^{,\,\mu})\,(\Phi^{c}_{,\,\nu}n_{c}\Phi^{d\,,\,\nu}n_{d})$ :
\begin{align}
 \alpha_{14}=&\; s^3
   \Bigg(\frac{\left(U'\right)^{10}}{2 G \varphi ^2 U^6}-\frac{9
   \left(U'\right)^{10}}{8 U^7}+\frac{9 \left(U'\right)^9}{8
   \varphi  U^6}+\frac{3 G' \left(U'\right)^8}{4 G \varphi 
   U^5}-\frac{19 U'' \left(U'\right)^8}{12 G \varphi ^2
   U^5}+\frac{9 U'' \left(U'\right)^8}{2 U^6}+\frac{29
   \left(U'\right)^8}{12 \varphi ^2 U^5}-\frac{G'
   \left(U'\right)^7}{3 G \varphi ^2 U^4}+\frac{9 G'
   \left(U'\right)^7}{4 U^5}\nonumber\\
&\;-\frac{9 U'' \left(U'\right)^7}{2
   \varphi  U^5}-\frac{5 \left(G'\right)^2 \left(U'\right)^6}{24
   G U^4}-\frac{155 \left(U''\right)^2 \left(U'\right)^6}{24 G
   \varphi ^2 U^4}-\frac{9 \left(U''\right)^2
   \left(U'\right)^6}{2 U^5}-\frac{19 G' \left(U'\right)^6}{8
   \varphi  U^4}-\frac{9 G'' \left(U'\right)^6}{8 U^4}-\frac{G'
   U'' \left(U'\right)^6}{G \varphi  U^4}\nonumber\\
&\;-\frac{343 U''
   \left(U'\right)^6}{36 \varphi ^2 U^4}-\frac{\left(G'\right)^2
   \left(U'\right)^5}{2 G \varphi  U^3}+\frac{9
   \left(U''\right)^2 \left(U'\right)^5}{2 \varphi 
   U^4}-\frac{29 G' \left(U'\right)^5}{18 \varphi ^2
   U^3}-\frac{5 G' U'' \left(U'\right)^5}{4 G \varphi ^2
   U^3}-\frac{15 G' U'' \left(U'\right)^5}{2 U^4}-\frac{95
   \left(G'\right)^2 \left(U'\right)^4}{72 U^3}\nonumber\\
&\;-\frac{31 G'
   \left(U''\right)^2 \left(U'\right)^4}{4 G \varphi 
   U^3}+\frac{493 \left(U''\right)^2 \left(U'\right)^4}{72
   \varphi ^2 U^3}-\frac{\left(G'\right)^2 U''
   \left(U'\right)^4}{3 G U^3}+\frac{26 G' U''
   \left(U'\right)^4}{3 \varphi  U^3}+\frac{9 G'' U''
   \left(U'\right)^4}{2 U^3}+\frac{29 \left(G'\right)^2
   \left(U'\right)^3}{24 \varphi  U^2}\nonumber\\
&\;+\frac{6 G'
   \left(U''\right)^2 \left(U'\right)^3}{U^3}+\frac{3 G' G''
   \left(U'\right)^3}{4 U^2}-\frac{3 \left(G'\right)^2 U''
   \left(U'\right)^3}{G \varphi  U^2}+\frac{31 G' U''
   \left(U'\right)^3}{12 \varphi ^2 U^2}-\frac{\left(G'\right)^3
   \left(U'\right)^2}{4 G \varphi  U}+\frac{\left(G'\right)^2
   \left(U'\right)^2}{4 \varphi ^2 U}\nonumber\\
&\;+\frac{31 \left(G'\right)^2
   \left(U''\right)^2 \left(U'\right)^2}{24 G U^2}-\frac{121 G'
   \left(U''\right)^2 \left(U'\right)^2}{12 \varphi 
   U^2}-\frac{9 G'' \left(U''\right)^2 \left(U'\right)^2}{2
   U^2}+\frac{43 \left(G'\right)^2 U'' \left(U'\right)^2}{18
   U^2}+\frac{\left(G'\right)^3 U'}{4 U}+\frac{\left(G'\right)^3
   U'' U'}{2 G U}\nonumber\\
&\;-\frac{7 \left(G'\right)^2 U'' U'}{2 \varphi 
   U}-\frac{3 G' G'' U'' U'}{2 U}+\frac{\left(G'\right)^4}{24
   G}-\frac{7 \left(G'\right)^3}{24 \varphi }-\frac{5
   \left(G'\right)^2 \left(U''\right)^2}{72 U}-\frac{1}{8}
   \left(G'\right)^2 G''\Bigg)\nonumber\\
&+s^2 \Bigg(-\frac{25 \left(U'\right)^8}{24 G^2
   \varphi ^4 U^4}-\frac{13 \left(U'\right)^8}{9 G \varphi ^2
   U^5}-\frac{15 \left(U'\right)^8}{2
   U^6}+\frac{\left(U'\right)^7}{2 G \varphi ^3 U^4}+\frac{39
   \left(U'\right)^7}{4 \varphi  U^5}-\frac{5 G'
   \left(U'\right)^6}{2 G^2 \varphi ^3 U^3}+\frac{7 G'
   \left(U'\right)^6}{12 G \varphi  U^4}-\frac{7 U''
   \left(U'\right)^6}{9 G \varphi ^2 U^4}\nonumber\\
&\;+\frac{141 U''
   \left(U'\right)^6}{8 U^5}+\frac{11 \left(U'\right)^6}{36 G
   \varphi ^4 U^3}-\frac{185 \left(U'\right)^6}{54 \varphi ^2
   U^4}+\frac{127 G' \left(U'\right)^5}{72 G \varphi ^2
   U^3}+\frac{11 G' \left(U'\right)^5}{8 U^4}-\frac{5 G''
   \left(U'\right)^5}{12 G \varphi  U^3}+\frac{5 G' U''
   \left(U'\right)^5}{4 G^2 \varphi ^2 U^3}+\frac{3 U''
   \left(U'\right)^5}{2 G \varphi ^3 U^3}\nonumber\\
&\;-\frac{111 U''
   \left(U'\right)^5}{4 \varphi  U^4}-\frac{3 U'''
   \left(U'\right)^5}{4 G \varphi ^2 U^3}-\frac{35
   \left(U'\right)^5}{6 \varphi ^3 U^3}-\frac{7
   \left(G'\right)^2 \left(U'\right)^4}{8 G^2 \varphi ^2
   U^2}+\frac{17 \left(G'\right)^2 \left(U'\right)^4}{18 G
   U^3}+\frac{55 \left(U''\right)^2 \left(U'\right)^4}{18 G
   \varphi ^2 U^3}-\frac{27 \left(U''\right)^2
   \left(U'\right)^4}{4 U^4}\nonumber\\
&\;+\frac{2 G' \left(U'\right)^4}{G
   \varphi ^3 U^2}-\frac{47 G' \left(U'\right)^4}{9 \varphi 
   U^3}+\frac{G'' \left(U'\right)^4}{U^3}-\frac{131 G' U''
   \left(U'\right)^4}{12 G \varphi  U^3}+\frac{1249 U''
   \left(U'\right)^4}{108 \varphi ^2 U^3}+\frac{695
   \left(U'\right)^4}{216 \varphi ^4
   U^2}-\frac{\left(G'\right)^2 \left(U'\right)^3}{4 G \varphi 
   U^2}\nonumber\\
&\;+\frac{39 \left(U''\right)^2 \left(U'\right)^3}{2 \varphi
    U^3}+\frac{1307 G' \left(U'\right)^3}{216 \varphi ^2
   U^2}+\frac{31 G'' \left(U'\right)^3}{36 \varphi  U^2}+\frac{3
   \left(G'\right)^2 U'' \left(U'\right)^3}{2 G^2 \varphi 
   U^2}-\frac{11 G' U'' \left(U'\right)^3}{12 G \varphi ^2
   U^2}+\frac{29 G' U'' \left(U'\right)^3}{4 U^3}\nonumber\\
&\;-\frac{13 G''
   U'' \left(U'\right)^3}{12 G \varphi  U^2}+\frac{U''
   \left(U'\right)^3}{2 \varphi ^3 U^2}+\frac{3 G' U'''
   \left(U'\right)^3}{4 G \varphi  U^2}-\frac{U'''
   \left(U'\right)^3}{4 \varphi ^2 U^2}+\frac{3
   \left(G'\right)^3 \left(U'\right)^2}{4 G^2 \varphi 
   U}-\frac{97 \left(G'\right)^2 \left(U'\right)^2}{24 G \varphi
   ^2 U}+\frac{2 \left(G'\right)^2 \left(U'\right)^2}{3
   U^2}\nonumber\\
&\;+\frac{25 G' \left(U''\right)^2 \left(U'\right)^2}{4 G
   \varphi  U^2}-\frac{583 \left(U''\right)^2
   \left(U'\right)^2}{216 \varphi ^2 U^2}-\frac{19 G'
   \left(U'\right)^2}{18 \varphi ^3 U}-\frac{7 G' G''
   \left(U'\right)^2}{4 G \varphi  U}+\frac{143
   \left(G'\right)^2 U'' \left(U'\right)^2}{72 G U^2}-\frac{6 G'
   U'' \left(U'\right)^2}{\varphi  U^2}\nonumber\\
&\;-\frac{17 G'' U''
   \left(U'\right)^2}{4 U^2}-\frac{11 \left(G'\right)^3 U'}{24 G
   U}-\frac{2 \left(G'\right)^2 U'}{9 \varphi  U}-\frac{13 G'
   \left(U''\right)^2 U'}{2 U^2}+\frac{G' G'' U'}{2
   U}-\frac{\left(G'\right)^3 U'' U'}{4 G^2 U}+\frac{2
   \left(G'\right)^2 U'' U'}{G \varphi  U}\nonumber\\
&\;+\frac{5 G' U'' U'}{12
   \varphi ^2 U}+\frac{G' G'' U'' U'}{2 G U}+\frac{23 G'' U''
   U'}{36 \varphi  U}+\frac{G' U''' U'}{4 \varphi 
   U}-\frac{\left(G'\right)^4}{12 G^2}+\frac{5
   \left(G'\right)^3}{6 G \varphi }-\frac{3 \left(G'\right)^2}{4
   \varphi ^2}-\frac{\left(G''\right)^2}{4}-\frac{49
   \left(G'\right)^2 \left(U''\right)^2}{72 G U}\nonumber\\
&\;+\frac{49 G'
   \left(U''\right)^2}{18 \varphi  U}+\frac{3 G''
   \left(U''\right)^2}{2 U}+\frac{\left(G'\right)^2 G''}{3
   G}-\frac{11 G' G''}{12 \varphi }-\frac{2 \left(G'\right)^2
   U''}{3 U}\Bigg)\nonumber\\
&+s \Bigg(\frac{25
   \left(U'\right)^6}{36 G^2 \varphi ^4 U^3}+\frac{43
   \left(U'\right)^6}{54 G \varphi ^2 U^4}+\frac{12
   \left(U'\right)^6}{U^5}+\frac{5 G' \left(U'\right)^5}{24 G^2
   \varphi ^2 U^3}-\frac{\left(U'\right)^5}{3 G \varphi ^3
   U^3}-\frac{7 \left(U'\right)^5}{\varphi  U^4}-\frac{5
   \left(G'\right)^2 \left(U'\right)^4}{24 G^3 \varphi ^2
   U^2}-\frac{5 G' \left(U'\right)^4}{24 G^2 \varphi ^3
   U^2}\nonumber\\
&\;-\frac{47 G' \left(U'\right)^4}{36 G \varphi 
   U^3}+\frac{5 G'' \left(U'\right)^4}{8 G^2 \varphi ^2
   U^2}+\frac{79 U'' \left(U'\right)^4}{54 G \varphi ^2
   U^3}-\frac{157 U'' \left(U'\right)^4}{8 U^4}+\frac{43
   \left(U'\right)^4}{108 G \varphi ^4 U^2}-\frac{3281
   \left(U'\right)^4}{324 \varphi ^2
   U^3}-\frac{\left(G'\right)^2 \left(U'\right)^3}{G^2 \varphi 
   U^2}+\frac{137 G' \left(U'\right)^3}{108 G \varphi ^2
   U^2}\nonumber\\
&\;-\frac{4 G' \left(U'\right)^3}{3 U^3}+\frac{41 G''
   \left(U'\right)^3}{36 G \varphi  U^2}-\frac{5 G' U''
   \left(U'\right)^3}{6 G^2 \varphi ^2 U^2}-\frac{U''
   \left(U'\right)^3}{G \varphi ^3 U^2}+\frac{23 U''
   \left(U'\right)^3}{\varphi  U^3}+\frac{U'''
   \left(U'\right)^3}{2 G \varphi ^2 U^2}+\frac{3 U'''
   \left(U'\right)^3}{2 U^3}+\frac{107 \left(U'\right)^3}{18
   \varphi ^3 U^2}\nonumber\\
&\;-\frac{\left(G'\right)^3 \left(U'\right)^2}{4
   G^3 \varphi  U}-\frac{25 \left(G'\right)^2
   \left(U'\right)^2}{9 G^2 \varphi ^2 U}-\frac{11
   \left(G'\right)^2 \left(U'\right)^2}{12 G U^2}-\frac{65
   \left(U''\right)^2 \left(U'\right)^2}{216 G \varphi ^2
   U^2}+\frac{15 \left(U''\right)^2 \left(U'\right)^2}{4
   U^3}-\frac{139 G' \left(U'\right)^2}{24 G \varphi ^3
   U}+\frac{1231 G' \left(U'\right)^2}{216 \varphi 
   U^2}\nonumber\\
&\;+\frac{G' G'' \left(U'\right)^2}{2 G^2 \varphi 
   U}+\frac{41 G'' \left(U'\right)^2}{24 G \varphi ^2
   U}-\frac{G'' \left(U'\right)^2}{24 U^2}+\frac{35 G' U''
   \left(U'\right)^2}{6 G \varphi  U^2}+\frac{145 U''
   \left(U'\right)^2}{108 \varphi ^2 U^2}-\frac{3 U'''
   \left(U'\right)^2}{\varphi  U^2}+\frac{\left(U'\right)^2}{6
   \varphi ^4 U}-\frac{\left(G'\right)^3 U'}{8 G^2 U}\nonumber\\
&\;-\frac{79
   \left(G'\right)^2 U'}{36 G \varphi  U}-\frac{11
   \left(U''\right)^2 U'}{2 \varphi  U^2}-\frac{41 G' U'}{12
   \varphi ^2 U}+\frac{G' G'' U'}{4 G U}+\frac{7 G'' U'}{6
   \varphi  U}-\frac{\left(G'\right)^2 U'' U'}{2 G^2 \varphi 
   U}+\frac{4 G' U'' U'}{9 G \varphi ^2 U}-\frac{2 G' U''
   U'}{U^2}\nonumber\\
&\;+\frac{13 G'' U'' U'}{36 G \varphi  U}-\frac{U''
   U'}{6 \varphi ^3 U}-\frac{G' U''' U'}{4 G \varphi 
   U}+\frac{U''' U'}{12 \varphi ^2
   U}+\frac{\left(G'\right)^4}{24 G^3}+\frac{11
   \left(G'\right)^3}{24 G^2 \varphi }-\frac{19
   \left(G'\right)^2}{72 U}+\frac{3 \left(G'\right)^2}{2 G
   \varphi ^2}-\frac{\left(G''\right)^2}{12 G}-\frac{11 G'
   \left(U''\right)^2}{9 G \varphi  U}\nonumber\\
&\;-\frac{\left(G'\right)^2
   G''}{24 G^2}-\frac{G' G''}{12 G \varphi
   }-\frac{\left(G'\right)^2 U''}{2 G U}+\frac{7 G' U''}{36
   \varphi  U}+\frac{2 G'' U''}{3 U}+\frac{G' U'''}{2
   U}\Bigg)\nonumber\\
&-\frac{25 \left(U'\right)^4}{216 G^2 \varphi ^4 U^2}-\frac{10
   \left(U'\right)^4}{81 G \varphi ^2 U^3}-\frac{25
   \left(U'\right)^4}{8 U^4}-\frac{5 G' \left(U'\right)^3}{72
   G^2 \varphi ^2 U^2}-\frac{53 \left(U'\right)^3}{18 G \varphi
   ^3 U^2}-\frac{6 \left(U'\right)^3}{\varphi  U^3}+\frac{5
   \left(G'\right)^2 \left(U'\right)^2}{72 G^3 \varphi ^2
   U}+\frac{25 G' \left(U'\right)^2}{72 G^2 \varphi ^3
   U}\nonumber\\
&\;+\frac{55 G' \left(U'\right)^2}{108 G \varphi  U^2}-\frac{5
   G'' \left(U'\right)^2}{24 G^2 \varphi ^2 U}+\frac{125 U''
   \left(U'\right)^2}{108 G \varphi ^2 U^2}+\frac{53 U''
   \left(U'\right)^2}{12 U^3}-\frac{\left(U'\right)^2}{6 G
   \varphi ^4 U}-\frac{27 \left(U'\right)^2}{2 \varphi ^2
   U^2}-\frac{3 G
   \left(U'\right)^2}{U^3}+\frac{\left(G'\right)^2 U'}{2 G^2
   \varphi  U}+\frac{35 G' U'}{12 G \varphi ^2 U}\nonumber\\
&\;+\frac{15 G'
   U'}{4 U^2}-\frac{G'' U'}{2 G \varphi  U}+\frac{5 G' U''
   U'}{36 G^2 \varphi ^2 U}+\frac{U'' U'}{6 G \varphi ^3
   U}+\frac{83 U'' U'}{6 \varphi  U^2}-\frac{U''' U'}{12 G
   \varphi ^2 U}-\frac{U''' U'}{2 U^2}-\frac{2 G U'}{\varphi 
   U^2}+\frac{\left(G'\right)^3}{12 G^3 \varphi }+\frac{55
   \left(G'\right)^2}{72 G U}\nonumber\\
&\;+\frac{3 \left(G'\right)^2}{4 G^2 \varphi
   ^2}-\frac{\left(U''\right)^2}{3 U^2}+\frac{G'}{\varphi 
   U}-\frac{G' G''}{6 G^2
   \varphi }-\frac{G''}{U}-\frac{37 G' U''}{36 G \varphi 
   U}+\frac{2 G U''}{U^2}-\frac{(N-1) \left(G'\right)^2}{2
   G^2 \varphi ^2}-\frac{(N-1) \left(G'\right)^3}{4 G^3 \varphi
   }+\frac{(N-1) G' G''}{2 G^2 \varphi }
\end{align}
\\

The coefficient $ \alpha_{15}$ in front of the structure $\Phi^{a}_{,\,\mu}\Phi_{a\,,\,\nu}\Phi^{b\,,\,\mu}\Phi_{b}^{,\,\nu}$ :
\begin{align}
 \alpha_{15}=&\; s^2 \Bigg(\frac{25 \left(U'\right)^8}{24
   G^2 \varphi ^4 U^4}+\frac{3 \left(U'\right)^8}{8 U^6}-\frac{3
   \left(U'\right)^7}{\varphi  U^5}+\frac{5 G'
   \left(U'\right)^6}{2 G^2 \varphi ^3 U^3}+\frac{205
   \left(U'\right)^6}{36 G \varphi ^4 U^3}+\frac{9
   \left(U'\right)^6}{\varphi ^2 U^4}+\frac{5 G'
   \left(U'\right)^5}{6 G \varphi ^2 U^3}+\frac{G'
   \left(U'\right)^5}{2 U^4}-\frac{12 \left(U'\right)^5}{\varphi
   ^3 U^3}\nonumber\\
&\;+\frac{13 \left(G'\right)^2 \left(U'\right)^4}{12 G^2
   \varphi ^2 U^2}+\frac{6 G' \left(U'\right)^4}{G \varphi ^3
   U^2}-\frac{3 G' \left(U'\right)^4}{\varphi  U^3}+\frac{1681
   \left(U'\right)^4}{216 \varphi ^4
   U^2}+\frac{\left(G'\right)^2 \left(U'\right)^3}{G \varphi 
   U^2}+\frac{113 G' \left(U'\right)^3}{18 \varphi ^2
   U^2}-\frac{\left(G'\right)^3 \left(U'\right)^2}{2 G^2 \varphi
    U}\nonumber\\
&\;-\frac{77 \left(G'\right)^2 \left(U'\right)^2}{36 G
   \varphi ^2 U}+\frac{\left(G'\right)^2 \left(U'\right)^2}{4
   U^2}-\frac{41 G' \left(U'\right)^2}{18 \varphi ^3
   U}-\frac{\left(G'\right)^3 U'}{6 G U}-\frac{2
   \left(G'\right)^2 U'}{3 \varphi 
   U}+\frac{\left(G'\right)^4}{24
   G^2}+\frac{\left(G'\right)^3}{6 G \varphi
   }+\frac{\left(G'\right)^2}{6 \varphi ^2}\Bigg)\nonumber\\
&+s \Bigg(-\frac{25
   \left(U'\right)^6}{36 G^2 \varphi ^4 U^3}+\frac{2
   \left(U'\right)^6}{U^5}-\frac{5 \left(U'\right)^5}{2 G
   \varphi ^3 U^3}+\frac{\left(U'\right)^5}{2 \varphi 
   U^4}-\frac{5 G' \left(U'\right)^4}{3 G^2 \varphi ^3
   U^2}-\frac{205 \left(U'\right)^4}{108 G \varphi ^4
   U^2}+\frac{7 \left(U'\right)^4}{\varphi ^2 U^3}-\frac{16 G'
   \left(U'\right)^3}{9 G \varphi ^2 U^2}\nonumber\\
&\;-\frac{G'
   \left(U'\right)^3}{6 U^3}+\frac{19 \left(U'\right)^3}{6
   \varphi ^3 U^2}-\frac{31 \left(G'\right)^2
   \left(U'\right)^2}{36 G^2 \varphi ^2 U}-\frac{2 G'
   \left(U'\right)^2}{G \varphi ^3 U}-\frac{5 G'
   \left(U'\right)^2}{2 \varphi  U^2}-\frac{\left(G'\right)^2
   U'}{3 G \varphi  U}-\frac{11 G' U'}{6 \varphi ^2
   U}+\frac{\left(G'\right)^3}{6 G^2 \varphi
   }+\frac{\left(G'\right)^2}{4 U}+\frac{\left(G'\right)^2}{3 G
   \varphi ^2}\Bigg)\nonumber\\
&+\frac{25 \left(U'\right)^4}{216 G^2 \varphi ^4 U^2}-\frac{17
   \left(U'\right)^4}{24 U^4}+\frac{2 \left(U'\right)^3}{3 G
   \varphi ^3 U^2}+\frac{\left(U'\right)^3}{6 \varphi 
   U^3}+\frac{G' \left(U'\right)^2}{9 G^2 \varphi ^3
   U}-\frac{\left(U'\right)^2}{3 G \varphi ^4 U}-\frac{5
   \left(U'\right)^2}{3 \varphi ^2 U^2}+\frac{G
   \left(U'\right)^2}{4 U^3}+\frac{G' U'}{2 G \varphi ^2
   U}+\frac{U'' U'}{3 G \varphi ^3 U}\nonumber\\
&\;+\frac{5 U'}{3 \varphi ^3
   U}-\frac{G U'}{6 \varphi  U^2}+\frac{\left(G'\right)^2}{6 G^2
   \varphi ^2}+\frac{7 G'}{6 \varphi  U}-\frac{5
   U''}{3 \varphi ^2 U}+\frac{3 G^2}{2 U^2}
\end{align}
\\

The coefficient $ \alpha_{16}$ in front of the structure $\Phi^{a}_{,\mu}n_{a}\Phi^{b}_{,\nu}n_{b}\Phi^{c\;,\,\mu}\Phi_{c}^{,\,\nu}$ :
\begin{align}
 \alpha_{16}=&\; s^3
   \Bigg(\frac{\left(U'\right)^{10}}{G \varphi ^2 U^6}-\frac{9
   \left(U'\right)^{10}}{4 U^7}+\frac{9
   \left(U'\right)^9}{\varphi  U^6}+\frac{3 G'
   \left(U'\right)^8}{2 G \varphi  U^5}-\frac{19 U''
   \left(U'\right)^8}{6 G \varphi ^2 U^5}+\frac{9 U''
   \left(U'\right)^8}{U^6}-\frac{26 \left(U'\right)^8}{3 \varphi
   ^2 U^5}-\frac{2 G' \left(U'\right)^7}{3 G \varphi ^2
   U^4}-\frac{36 U'' \left(U'\right)^7}{\varphi  U^5}\nonumber\\
&\;-\frac{5
   \left(G'\right)^2 \left(U'\right)^6}{12 G U^4}-\frac{155
   \left(U''\right)^2 \left(U'\right)^6}{12 G \varphi ^2
   U^4}-\frac{9 \left(U''\right)^2
   \left(U'\right)^6}{U^5}-\frac{5 G' \left(U'\right)^6}{2
   \varphi  U^4}-\frac{2 G' U'' \left(U'\right)^6}{G \varphi 
   U^4}+\frac{629 U'' \left(U'\right)^6}{18 \varphi ^2
   U^4}-\frac{\left(G'\right)^2 \left(U'\right)^5}{G \varphi 
   U^3}\nonumber\\
&\;+\frac{36 \left(U''\right)^2 \left(U'\right)^5}{\varphi 
   U^4}+\frac{52 G' \left(U'\right)^5}{9 \varphi ^2 U^3}-\frac{5
   G' U'' \left(U'\right)^5}{2 G \varphi ^2 U^3}+\frac{3 G' U''
   \left(U'\right)^5}{U^4}+\frac{13 \left(G'\right)^2
   \left(U'\right)^4}{36 U^3}-\frac{31 G' \left(U''\right)^2
   \left(U'\right)^4}{2 G \varphi  U^3}\nonumber\\
&\;-\frac{1451
   \left(U''\right)^2 \left(U'\right)^4}{36 \varphi ^2
   U^3}-\frac{2 \left(G'\right)^2 U'' \left(U'\right)^4}{3 G
   U^3}-\frac{2 G' U'' \left(U'\right)^4}{3 \varphi 
   U^3}-\frac{4 \left(G'\right)^2 \left(U'\right)^3}{3 \varphi 
   U^2}-\frac{6 G' \left(U''\right)^2
   \left(U'\right)^3}{U^3}-\frac{6 \left(G'\right)^2 U''
   \left(U'\right)^3}{G \varphi  U^2}\nonumber\\
&\;-\frac{77 G' U''
   \left(U'\right)^3}{6 \varphi ^2 U^2}-\frac{\left(G'\right)^3
   \left(U'\right)^2}{2 G \varphi  U}-\frac{\left(G'\right)^2
   \left(U'\right)^2}{\varphi ^2 U}+\frac{31 \left(G'\right)^2
   \left(U''\right)^2 \left(U'\right)^2}{12 G U^2}+\frac{41 G'
   \left(U''\right)^2 \left(U'\right)^2}{6 \varphi 
   U^2}-\frac{11 \left(G'\right)^2 U'' \left(U'\right)^2}{9
   U^2}\nonumber\\
&\;+\frac{\left(G'\right)^3 U'' U'}{G U}+\frac{2
   \left(G'\right)^2 U'' U'}{\varphi 
   U}+\frac{\left(G'\right)^4}{12 G}+\frac{\left(G'\right)^3}{6
   \varphi }-\frac{5 \left(G'\right)^2 \left(U''\right)^2}{36
   U}\Bigg)\nonumber\\
&+s^2 \Bigg(-\frac{25 \left(U'\right)^8}{12 G^2
   \varphi ^4 U^4}+\frac{83 \left(U'\right)^8}{18 G \varphi ^2
   U^5}+\frac{15 \left(U'\right)^8}{2
   U^6}+\frac{\left(U'\right)^7}{G \varphi ^3 U^4}+\frac{9
   \left(U'\right)^7}{\varphi  U^5}-\frac{5 G'
   \left(U'\right)^6}{G^2 \varphi ^3 U^3}-\frac{5 G'
   \left(U'\right)^6}{6 G \varphi  U^4}+\frac{17 U''
   \left(U'\right)^6}{18 G \varphi ^2 U^4}\nonumber\\
&\;-\frac{33 U''
   \left(U'\right)^6}{2 U^5}-\frac{205 \left(U'\right)^6}{18 G
   \varphi ^4 U^3}-\frac{73 \left(U'\right)^6}{54 \varphi ^2
   U^4}-\frac{17 G' \left(U'\right)^5}{36 G \varphi ^2
   U^3}-\frac{19 G' \left(U'\right)^5}{4 U^4}+\frac{5 G''
   \left(U'\right)^5}{3 G \varphi  U^3}+\frac{5 G' U''
   \left(U'\right)^5}{2 G^2 \varphi ^2 U^3}\nonumber\\
&\;+\frac{15 U''
   \left(U'\right)^5}{2 G \varphi ^3 U^3}+\frac{6 U''
   \left(U'\right)^5}{\varphi  U^4}+\frac{73
   \left(U'\right)^5}{3 \varphi ^3 U^3}-\frac{7
   \left(G'\right)^2 \left(U'\right)^4}{4 G^2 \varphi ^2
   U^2}+\frac{7 \left(G'\right)^2 \left(U'\right)^4}{18 G
   U^3}+\frac{55 \left(U''\right)^2 \left(U'\right)^4}{9 G
   \varphi ^2 U^3}-\frac{19 G' \left(U'\right)^4}{2 G \varphi ^3
   U^2}\nonumber\\
&\;-\frac{11 G' \left(U'\right)^4}{18 \varphi 
   U^3}-\frac{G'' \left(U'\right)^4}{U^3}-\frac{G' U''
   \left(U'\right)^4}{3 G \varphi  U^3}-\frac{973 U''
   \left(U'\right)^4}{27 \varphi ^2 U^3}-\frac{1681
   \left(U'\right)^4}{108 \varphi ^4 U^2}-\frac{24
   \left(U''\right)^2 \left(U'\right)^3}{\varphi 
   U^3}-\frac{1537 G' \left(U'\right)^3}{108 \varphi ^2
   U^2}\nonumber\\
&\;+\frac{23 G'' \left(U'\right)^3}{9 \varphi  U^2}+\frac{3
   \left(G'\right)^2 U'' \left(U'\right)^3}{G^2 \varphi 
   U^2}+\frac{47 G' U'' \left(U'\right)^3}{3 G \varphi ^2
   U^2}+\frac{G' U'' \left(U'\right)^3}{U^3}+\frac{13 G'' U''
   \left(U'\right)^3}{3 G \varphi  U^2}+\frac{5 U''
   \left(U'\right)^3}{2 \varphi ^3 U^2}+\frac{3
   \left(G'\right)^3 \left(U'\right)^2}{2 G^2 \varphi 
   U}\nonumber\\
&\;+\frac{83 \left(G'\right)^2 \left(U'\right)^2}{12 G \varphi
   ^2 U}+\frac{\left(G'\right)^2 \left(U'\right)^2}{3
   U^2}+\frac{9 G' \left(U''\right)^2 \left(U'\right)^2}{2 G
   \varphi  U^2}+\frac{1361 \left(U''\right)^2
   \left(U'\right)^2}{108 \varphi ^2 U^2}+\frac{97 G'
   \left(U'\right)^2}{18 \varphi ^3 U}+\frac{G' G''
   \left(U'\right)^2}{G \varphi  U}\nonumber\\
&\;+\frac{29 \left(G'\right)^2
   U'' \left(U'\right)^2}{9 G U^2}+\frac{20 G' U''
   \left(U'\right)^2}{3 \varphi  U^2}+\frac{2 G'' U''
   \left(U'\right)^2}{U^2}+\frac{7 \left(G'\right)^3 U'}{12 G
   U}+\frac{26 \left(G'\right)^2 U'}{9 \varphi  U}+\frac{2 G'
   \left(U''\right)^2 U'}{U^2}-\frac{\left(G'\right)^3 U'' U'}{2
   G^2 U}\nonumber\\
&\;-\frac{2 \left(G'\right)^2 U'' U'}{G \varphi 
   U}+\frac{2 G' U'' U'}{3 \varphi ^2 U}-\frac{2 G' G'' U''
   U'}{G U}-\frac{23 G'' U'' U'}{9 \varphi 
   U}-\frac{\left(G'\right)^4}{6 G^2}-\frac{5
   \left(G'\right)^3}{6 G \varphi }-\frac{\left(G'\right)^2}{2
   \varphi ^2}+\frac{5 \left(G'\right)^2 \left(U''\right)^2}{36
   G U}\nonumber\\
&-\frac{2 G' \left(U''\right)^2}{9 \varphi 
   U}-\frac{\left(G'\right)^2 G''}{3 G}-\frac{G' G''}{3 \varphi
   }+\frac{2 \left(G'\right)^2 U''}{3 U}\Bigg)\nonumber\\
&+s \Bigg(\frac{25 \left(U'\right)^6}{18 G^2 \varphi ^4
   U^3}+\frac{167 \left(U'\right)^6}{54 G \varphi ^2
   U^4}-\frac{9 \left(U'\right)^6}{U^5}-\frac{13 G'
   \left(U'\right)^5}{12 G^2 \varphi ^2 U^3}+\frac{53
   \left(U'\right)^5}{6 G \varphi ^3 U^3}-\frac{25
   \left(U'\right)^5}{2 \varphi  U^4}-\frac{5 \left(G'\right)^2
   \left(U'\right)^4}{12 G^3 \varphi ^2 U^2}+\frac{11 G'
   \left(U'\right)^4}{6 G^2 \varphi ^3 U^2}\nonumber\\
&\;+\frac{20 G'
   \left(U'\right)^4}{9 G \varphi  U^3}-\frac{175 U''
   \left(U'\right)^4}{54 G \varphi ^2 U^3}+\frac{13 U''
   \left(U'\right)^4}{2 U^4}+\frac{151 \left(U'\right)^4}{54 G
   \varphi ^4 U^2}-\frac{1735 \left(U'\right)^4}{81 \varphi ^2
   U^3}+\frac{7 \left(G'\right)^2 \left(U'\right)^3}{2 G^2
   \varphi  U^2}+\frac{172 G' \left(U'\right)^3}{27 G \varphi ^2
   U^2}+\frac{17 G' \left(U'\right)^3}{6 U^3}\nonumber\\
&\;+\frac{4 G''
   \left(U'\right)^3}{9 G \varphi  U^2}-\frac{G' U''
   \left(U'\right)^3}{6 G^2 \varphi ^2 U^2}-\frac{U''
   \left(U'\right)^3}{G \varphi ^3 U^2}+\frac{23 U''
   \left(U'\right)^3}{\varphi  U^3}-\frac{161
   \left(U'\right)^3}{18 \varphi ^3 U^2}-\frac{\left(G'\right)^3
   \left(U'\right)^2}{2 G^3 \varphi  U}+\frac{4
   \left(G'\right)^2 \left(U'\right)^2}{9 G^2 \varphi ^2
   U}+\frac{2 \left(G'\right)^2 \left(U'\right)^2}{3 G
   U^2}\nonumber\\
&\;-\frac{119 \left(U''\right)^2 \left(U'\right)^2}{108 G
   \varphi ^2 U^2}+\frac{3 \left(U''\right)^2
   \left(U'\right)^2}{U^3}+\frac{8 G' \left(U'\right)^2}{3 G
   \varphi ^3 U}+\frac{289 G' \left(U'\right)^2}{27 \varphi 
   U^2}-\frac{G' G'' \left(U'\right)^2}{2 G^2 \varphi 
   U}-\frac{G'' \left(U'\right)^2}{3 U^2}-\frac{3 G' U''
   \left(U'\right)^2}{G \varphi  U^2}\nonumber\\
&\;+\frac{373 U''
   \left(U'\right)^2}{54 \varphi ^2
   U^2}-\frac{\left(U'\right)^2}{3 \varphi ^4
   U}-\frac{\left(G'\right)^3 U'}{4 G^2 U}+\frac{5
   \left(G'\right)^2 U'}{18 G \varphi  U}+\frac{4
   \left(U''\right)^2 U'}{\varphi  U^2}+\frac{9 G' U'}{2 \varphi
   ^2 U}-\frac{G' G'' U'}{2 G U}-\frac{2 G'' U'}{3 \varphi 
   U}-\frac{\left(G'\right)^2 U'' U'}{G^2 \varphi  U}\nonumber\\
&\;-\frac{40
   G' U'' U'}{9 G \varphi ^2 U}-\frac{9 G' U'' U'}{2
   U^2}-\frac{13 G'' U'' U'}{9 G \varphi  U}+\frac{U'' U'}{2
   \varphi ^3 U}+\frac{\left(G'\right)^4}{12
   G^3}-\frac{\left(G'\right)^3}{3 G^2 \varphi }-\frac{25
   \left(G'\right)^2}{36 U}-\frac{\left(G'\right)^2}{2 G \varphi
   ^2}+\frac{\left(G''\right)^2}{3 G}+\frac{2 G'
   \left(U''\right)^2}{9 G \varphi 
   U}\nonumber\\
&\;-\frac{\left(U''\right)^2}{6 \varphi ^2
   U}+\frac{\left(G'\right)^2 G''}{6 G^2}+\frac{5 G' G''}{6 G
   \varphi }-\frac{25 G' U''}{9 \varphi  U}-\frac{2 G'' U''}{3
   U}\Bigg)\nonumber\\
&-\frac{25 \left(U'\right)^4}{108 G^2 \varphi ^4 U^2}-\frac{128
   \left(U'\right)^4}{81 G \varphi ^2 U^3}+\frac{9
   \left(U'\right)^4}{4 U^4}+\frac{13 G' \left(U'\right)^3}{36
   G^2 \varphi ^2 U^2}-\frac{55 \left(U'\right)^3}{18 G \varphi
   ^3 U^2}+\frac{7 \left(U'\right)^3}{2 \varphi  U^3}+\frac{5
   \left(G'\right)^2 \left(U'\right)^2}{36 G^3 \varphi ^2
   U}-\frac{G' \left(U'\right)^2}{18 G^2 \varphi ^3 U}+\frac{53
   G' \left(U'\right)^2}{27 G \varphi  U^2}\nonumber\\
&\;+\frac{59 U''
   \left(U'\right)^2}{54 G \varphi ^2 U^2}-\frac{2 U''
   \left(U'\right)^2}{3 U^3}+\frac{\left(U'\right)^2}{3 G
   \varphi ^4 U}+\frac{8 \left(U'\right)^2}{\varphi ^2
   U^2}+\frac{19 G \left(U'\right)^2}{2 U^3}-\frac{G' U'}{2 G
   \varphi ^2 U}-\frac{13 G' U'}{2 U^2}-\frac{2 G' U'' U'}{9 G^2
   \varphi ^2 U}-\frac{U'' U'}{2 G \varphi ^3 U}\nonumber\\
&\;-\frac{22 U''
   U'}{3 \varphi  U^2}+\frac{\left(G'\right)^3}{6 G^3 \varphi
   }+\frac{7 \left(G'\right)^2}{36 G
   U}+\frac{\left(U''\right)^2}{6 G \varphi ^2 U}-\frac{2
   \left(U''\right)^2}{3 U^2}-\frac{G'}{\varphi  U}+\frac{G'
   G''}{6 G^2 \varphi }+\frac{G''}{U}+\frac{G' U''}{9 G \varphi 
   U}
\end{align}
\\

The coefficient $ \alpha_{17}$ in front of the structure $\Phi^{a\;\;\mu}_{;\,\mu}\Phi_{a;\,\nu}^{\;\;\;\;\;\nu}$:
\begin{align}
 \alpha_{17}=&\; s \left(-\frac{\left(U'\right)^4}{4 G U^2 \varphi ^2}-\frac{G'
   U'}{2 U}+\frac{\left(G'\right)^2}{4
   G}-\frac{\left(U'\right)^2}{12 U \varphi ^2}-\frac{3
   \left(U'\right)^4}{4 U^3}\right)+\frac{\left(U'\right)^2}{12
   G U \varphi ^2}-\frac{U'}{U \varphi
   }+\frac{\left(U'\right)^2}{4 U^2}
\end{align}
\\

The coefficient $ \alpha_{18}$ in front of the structure $(\Phi^{c\;\;\mu}_{;\,\mu}\,n_{c})^2$ :
\begin{align}
 \alpha_{18}=&\; s^2
   \left(\frac{9 G' \left(U'\right)^3}{4
   U^2}+\frac{\left(G'\right)^2}{8}+\frac{81
   \left(U'\right)^6}{8 U^4}\right)+ s
   \left(\frac{\left(U'\right)^4}{4 G U^2 \varphi
   ^2}-\frac{\left(G'\right)^2}{4 G}+\frac{\left(U'\right)^2}{12
   U \varphi ^2}-\frac{27 \left(U'\right)^4}{4 U^3}\right)-\frac{\left(U'\right)^2}{12
   G U \varphi ^2}+\frac{U'}{U \varphi }+\frac{13
   \left(U'\right)^2}{2 U^2}\nonumber\\
&\;-\frac{U''}{U}+\frac{(N-1) \left(G'\right)^2}{8 G^2}
\end{align}
\\

The coefficient $ \alpha_{19}$ in front of the structure $(\Phi^{a}_{,\,\mu}\Phi_{a}^{,\,\mu})(\Phi^{b\;\;\nu}_{;\,\nu}n_{b})$ :
\begin{align}
 \alpha_{19}=&\; s^2 \Bigg(-\frac{\left(U'\right)^7}{4 G \varphi ^2
   U^4}-\frac{27 \left(U'\right)^7}{8 U^5}+\frac{27
   \left(U'\right)^6}{4 \varphi  U^4}-\frac{3 G'
   \left(U'\right)^5}{4 G \varphi  U^3}-\frac{3 U''
   \left(U'\right)^5}{4 G \varphi ^2
   U^3}-\frac{\left(U'\right)^5}{12 \varphi ^2 U^3}+\frac{15 G'
   \left(U'\right)^4}{8 U^3}+\frac{3 \left(G'\right)^2
   \left(U'\right)^3}{8 G U^2}\nonumber\\
&\;-\frac{7 G' \left(U'\right)^3}{4
   \varphi  U^2}-\frac{9 G'' \left(U'\right)^3}{4 U^2}+\frac{3
   G' U'' \left(U'\right)^3}{4 G \varphi  U^2}-\frac{U''
   \left(U'\right)^3}{4 \varphi ^2 U^2}-\frac{3
   \left(G'\right)^2 \left(U'\right)^2}{4 G \varphi  U}+\frac{G'
   U'' U'}{4 \varphi  U}+\frac{\left(G'\right)^3}{8
   G}-\frac{\left(G'\right)^2}{2 \varphi }-\frac{G'
   G''}{4}\Bigg)\nonumber\\
&+s
   \Bigg(\frac{\left(U'\right)^5}{6 G \varphi ^2 U^3}+\frac{15
   \left(U'\right)^5}{8 U^4}+\frac{5 G' \left(U'\right)^4}{8 G^2
   \varphi ^2 U^2}-\frac{2 \left(U'\right)^4}{\varphi 
   U^3}+\frac{3 G' \left(U'\right)^3}{4 G \varphi 
   U^2}+\frac{U'' \left(U'\right)^3}{2 G \varphi ^2 U^2}+\frac{3
   U'' \left(U'\right)^3}{2 U^3}+\frac{37 \left(U'\right)^3}{36
   \varphi ^2 U^2}+\frac{3 \left(G'\right)^2
   \left(U'\right)^2}{4 G^2 \varphi  U}\nonumber\\
&\;+\frac{41 G'
   \left(U'\right)^2}{24 G \varphi ^2 U}-\frac{3 G'
   \left(U'\right)^2}{2 U^2}-\frac{3 U''
   \left(U'\right)^2}{\varphi  U^2}+\frac{\left(G'\right)^2
   U'}{4 G U}+\frac{G' U'}{\varphi  U}+\frac{G'' U'}{2
   U}-\frac{G' U'' U'}{4 G \varphi  U}+\frac{U'' U'}{12 \varphi
   ^2 U}-\frac{\left(G'\right)^3}{8 G^2}+\frac{G' U''}{2
   U}\Bigg)\nonumber\\
&+\frac{53 \left(U'\right)^3}{36 G \varphi ^2
   U^2}-\frac{\left(U'\right)^3}{4 U^3}-\frac{5 G'
   \left(U'\right)^2}{24 G^2 \varphi ^2 U}+\frac{14
   \left(U'\right)^2}{\varphi  U^2}-\frac{G' U'}{2 G \varphi 
   U}-\frac{U'' U'}{12 G \varphi ^2 U}-\frac{U'' U'}{2
   U^2}+\frac{U'}{\varphi ^2 U}+\frac{2 G U'}{U^2}-\frac{\left(G'\right)^2}{4
   G^2 \varphi }-\frac{G'}{2 U}-\frac{U''}{\varphi  U}\nonumber\\
&\;+\frac{(N-1)
   \left(G'\right)^2}{2 G^2 \varphi }
\end{align}
\\

The coefficient $ \alpha_{20}$ in front of the structure $(\Phi^{a}_{,\,\mu}n_{a}\Phi^{b\,,\,\mu}n_{b})(\Phi^{c\;\;\nu}_{;\,\nu}\,n_{c})$ :
\begin{align}
 \alpha_{20}=&\; s^3\Bigg(\frac{81 \left(U'\right)^9}{8 U^6}-\frac{81 U''
   \left(U'\right)^7}{2 U^5}-\frac{45 G' \left(U'\right)^6}{8
   U^4}+\frac{81 \left(U''\right)^2 \left(U'\right)^5}{2
   U^4}+\frac{9 G' U'' \left(U'\right)^4}{U^3}+\frac{3
   \left(G'\right)^2 \left(U'\right)^3}{8 U^2}+\frac{9 G'
   \left(U''\right)^2 \left(U'\right)^2}{2 U^2}\nonumber\\
&\;+\frac{3
   \left(G'\right)^2 U'' U'}{2
   U}+\frac{\left(G'\right)^3}{8}\Bigg)+ s^2\Bigg(\frac{3
   \left(U'\right)^7}{4 G \varphi ^2 U^4}-\frac{45
   \left(U'\right)^7}{8 U^5}-\frac{45 \left(U'\right)^6}{4
   \varphi  U^4}-\frac{G' \left(U'\right)^5}{2 G \varphi 
   U^3}+\frac{9 U'' \left(U'\right)^5}{2 G \varphi ^2
   U^3}+\frac{135 U'' \left(U'\right)^5}{4
   U^4}\nonumber\\
&\;+\frac{\left(U'\right)^5}{4 \varphi ^2 U^3}+\frac{5 G'
   \left(U'\right)^4}{4 G \varphi ^2 U^2}-\frac{21 G'
   \left(U'\right)^4}{8 U^3}+\frac{9 U''
   \left(U'\right)^4}{\varphi  U^3}-\frac{3 \left(G'\right)^2
   \left(U'\right)^3}{8 G U^2}-\frac{63 \left(U''\right)^2
   \left(U'\right)^3}{2 U^3}+\frac{4 G' \left(U'\right)^3}{3
   \varphi  U^2}+\frac{9 G'' \left(U'\right)^3}{4 U^2}\nonumber\\
&\;-\frac{13
   G' U'' \left(U'\right)^3}{4 G \varphi  U^2}+\frac{3 U''
   \left(U'\right)^3}{2 \varphi ^2 U^2}+\frac{5 G'
   \left(U'\right)^2}{12 \varphi ^2 U}-\frac{15 G' U''
   \left(U'\right)^2}{4 U^2}-\frac{\left(G'\right)^2 U'}{2
   U}+\frac{3 \left(G'\right)^2 U'' U'}{2 G U}+\frac{23 G' U''
   U'}{12 \varphi  U}+\frac{\left(G'\right)^3}{8 G}\nonumber\\
&\;+\frac{3
   \left(G'\right)^2}{4 \varphi }-\frac{3 G'
   \left(U''\right)^2}{2 U}+\frac{G' G''}{4}\Bigg)+s\Bigg(\frac{7 \left(U'\right)^5}{4 G \varphi ^2
   U^3}+\frac{15 \left(U'\right)^5}{8 U^4}-\frac{11 G'
   \left(U'\right)^4}{8 G^2 \varphi ^2
   U^2}-\frac{\left(U'\right)^4}{G \varphi ^3 U^2}+\frac{3
   \left(U'\right)^4}{\varphi  U^3}-\frac{25 G'
   \left(U'\right)^3}{12 G \varphi  U^2}\nonumber\\
 & -\frac{U''
   \left(U'\right)^3}{G \varphi ^2 U^2}-\frac{12 U''
   \left(U'\right)^3}{U^3}-\frac{7 \left(U'\right)^3}{3 \varphi
   ^2 U^2}-\frac{19 G' \left(U'\right)^2}{8 G \varphi ^2
   U}+\frac{3 G' \left(U'\right)^2}{U^2}-\frac{3 U''
   \left(U'\right)^2}{\varphi  U^2}-\frac{\left(U'\right)^2}{3
   \varphi ^3 U}+\frac{\left(G'\right)^2 U'}{2 G U}+\frac{15
   \left(U''\right)^2 U'}{2 U^2}\nonumber\\
&\;-\frac{G' U'}{4 \varphi 
   U}+\frac{13 G' U'' U'}{12 G \varphi  U}+\frac{U'' U'}{6
   \varphi ^2 U}-\frac{\left(G'\right)^3}{8
   G^2}-\frac{\left(G'\right)^2}{2 G \varphi }-\frac{G' G''}{2
   G}\Bigg)-\frac{13
   \left(U'\right)^3}{6 G \varphi ^2 U^2}-\frac{29
   \left(U'\right)^3}{4 U^3}+\frac{11 G' \left(U'\right)^2}{24 G^2 \varphi
   ^2 U}+\frac{\left(U'\right)^2}{3 G \varphi ^3 U}\nonumber\\
&\;-\frac{33
   \left(U'\right)^2}{2 \varphi  U^2}+\frac{G' U'}{4 G \varphi 
   U}-\frac{3 U'}{\varphi ^2 U}-\frac{U' U''}{6 G \varphi ^2 U}+\frac{33 U' U''}{2
   U^2}+\frac{3 U''}{\varphi  U}-\frac{U'''}{U}-\frac{(N-1) \left(G'\right)^2}{4
   G^2 \varphi }-\frac{(N-1) \left(G'\right)^3}{8 G^3}\nonumber\\
&\;+\frac{(N-1) G' G''}{4
   G^2}
\end{align}
\\

The coefficient $ \alpha_{21}$ in front of the structure $\Phi^{a\;\;\mu}_{;\,\mu}\Phi_{a}^{,\,\nu}\Phi^{b}_{,\,\nu}n_{b}$ :
\begin{align}
 \alpha_{21}=&\; s^2
   \Bigg(-\frac{\left(U'\right)^7}{2 G \varphi ^2 U^4}-\frac{9
   \left(U'\right)^7}{4 U^5}+\frac{9 \left(U'\right)^6}{2
   \varphi  U^4}+\frac{5 G' \left(U'\right)^5}{4 G \varphi 
   U^3}-\frac{15 U'' \left(U'\right)^5}{4 G \varphi ^2
   U^3}+\frac{9 U'' \left(U'\right)^5}{2
   U^4}-\frac{\left(U'\right)^5}{6 \varphi ^2 U^3}-\frac{5 G'
   \left(U'\right)^4}{4 G \varphi ^2 U^2}-\frac{3 G'
   \left(U'\right)^4}{4 U^3}\nonumber\\
&\;-\frac{9 U''
   \left(U'\right)^4}{\varphi  U^3}+\frac{5 G'
   \left(U'\right)^3}{12 \varphi  U^2}+\frac{5 G' U''
   \left(U'\right)^3}{2 G \varphi  U^2}-\frac{5 U''
   \left(U'\right)^3}{4 \varphi ^2 U^2}+\frac{3
   \left(G'\right)^2 \left(U'\right)^2}{4 G \varphi  U}-\frac{5
   G' \left(U'\right)^2}{12 \varphi ^2 U}+\frac{3 G' U''
   \left(U'\right)^2}{U^2}+\frac{\left(G'\right)^2 U'}{4
   U}\nonumber\\
&\;-\frac{3 \left(G'\right)^2 U'' U'}{2 G U}-\frac{13 G' U''
   U'}{6 \varphi  U}-\frac{\left(G'\right)^3}{4
   G}-\frac{\left(G'\right)^2}{4 \varphi }\Bigg)\nonumber\\
&+s \Bigg(-\frac{23
   \left(U'\right)^5}{12 G \varphi ^2 U^3}+\frac{3
   \left(U'\right)^5}{4 U^4}+\frac{3 G' \left(U'\right)^4}{4 G^2
   \varphi ^2 U^2}+\frac{\left(U'\right)^4}{G \varphi ^3
   U^2}-\frac{\left(U'\right)^4}{\varphi  U^3}+\frac{4 G'
   \left(U'\right)^3}{3 G \varphi  U^2}+\frac{U''
   \left(U'\right)^3}{2 G \varphi ^2 U^2}+\frac{3 U''
   \left(U'\right)^3}{2 U^3}+\frac{47 \left(U'\right)^3}{36
   \varphi ^2 U^2}\nonumber\\
&\;-\frac{3 \left(G'\right)^2
   \left(U'\right)^2}{4 G^2 \varphi  U}+\frac{2 G'
   \left(U'\right)^2}{3 G \varphi ^2 U}-\frac{G'
   \left(U'\right)^2}{2 U^2}+\frac{6 U''
   \left(U'\right)^2}{\varphi  U^2}+\frac{\left(U'\right)^2}{3
   \varphi ^3 U}-\frac{3 \left(G'\right)^2 U'}{4 G U}-\frac{3 G'
   U'}{4 \varphi  U}-\frac{G'' U'}{2 U}-\frac{5 G' U'' U'}{6 G
   \varphi  U}\nonumber\\
&\;-\frac{U'' U'}{4 \varphi ^2
   U}+\frac{\left(G'\right)^3}{4 G^2}+\frac{\left(G'\right)^2}{2
   G \varphi }+\frac{G' G''}{2 G}\Bigg)+\frac{25 \left(U'\right)^3}{36 G \varphi ^2 U^2}-\frac{G'
   \left(U'\right)^2}{4 G^2 \varphi ^2
   U}-\frac{\left(U'\right)^2}{3 G \varphi ^3 U}+\frac{5
   \left(U'\right)^2}{2 \varphi  U^2}+\frac{G' U'}{4 G \varphi 
   U}+\frac{U'' U'}{4 G \varphi ^2 U}-\frac{U'' U'}{U^2}\nonumber\\
&\;+\frac{2
   U'}{\varphi ^2 U}-\frac{13 G U'}{2
   U^2}+\frac{\left(G'\right)^2}{4 G^2 \varphi
   }+\frac{G'}{U}-\frac{2 U''}{\varphi  U}
\end{align}
\\

\newpage
\twocolumngrid

\end{document}